\documentclass{article}

\usepackage{arxiv}

\usepackage[utf8]{inputenc} 
\usepackage[T1]{fontenc}    
\usepackage{xr-hyper}
\usepackage{hyperref}       
\usepackage{url}            
\usepackage{booktabs}       
\usepackage{amsfonts}       
\usepackage{nicefrac}       
\usepackage{microtype}      
\usepackage{lipsum}		
\usepackage{graphicx}
\usepackage[sectionbib,authoryear]{natbib}
\usepackage{doi}
\usepackage{amsmath, amssymb,amsthm,placeins}
\usepackage{setspace}
\usepackage{verbatim}
\usepackage{bbm}
\usepackage{xcolor}
\usepackage{dsfont}
\usepackage{multirow}
\usepackage{multicol}
\usepackage{floatrow}
\usepackage{setspace}
\usepackage{thmtools, thm-restate}
\usepackage{titlesec}

\newcommand{\bo}[1]{\boldsymbol{#1}}

\DeclareMathOperator{\vect}{vec}

\DeclareMathOperator{\argmax}{argmax}
\DeclareMathOperator{\Tr}{Tr}
\DeclareMathOperator{\Var}{Var}

\newcommand{\bm}[1]{\boldsymbol{#1}}

\title{Fast, universal estimation of latent variable models using extended variational approximations}

\author{Pekka~Korhonen\thanks{Corresponding author: Pekka Korhonen, \emph{email:} pekka.o.korhonen@jyu.fi}\\
	Department of Mathematics and Statistics\\
	University of Jyväskylä, Finland \\
	\And
	Francis K.C.~Hui \\
	Research School of Finance, \\
	Actuarial Studies \& Statistics, \\
	Australian National University, Australia \\
	\And
	Jenni~Niku \\
    Department of Mathematics and Statistics\\
	University of Jyväskylä, Finland \\
	\And
	Sara~Taskinen\\
	Department of Mathematics and Statistics\\
	University of Jyväskylä, Finland \and
}

\renewcommand{\shorttitle}{Fast, universal estimation of latent variable models}

\hypersetup{
pdftitle={Fast, universal estimation of latent variable models using extended variational approximations},
pdfsubject={stat.CO, stat.ME},
pdfauthor={Pekka~Korhonen, Francis K.C.~Hui, Jenni~Niku, Sara~Taskinen},
pdfkeywords={generalized linear latent variable models, Laplace approximation, multi-response data, multivariate abundance data, ordination, variational approximations},
}

\begin{document}
\maketitle

\begin{abstract}
	Generalized linear latent variable models (GLLVMs) are a class of methods for analyzing multi-response data which has garnered considerable popularity in recent years, for example, in the analysis of multivariate abundance data in ecology. One of the main features of GLLVMs is their capacity to handle a variety of responses types, such as (overdispersed) counts, binomial responses, (semi-)continuous, and proportions data. 
On the other hand, the introduction of underlying latent variables presents some major computational challenges, as the resulting marginal likelihood function involves an intractable integral for non-normally distributed responses. This has spurred research into approximation methods to overcome this integral, with a recent and particularly computationally scalable one being that of variational approximations (VA). However, research into the use of VA of GLLVMs and related models has been hampered by the fact that closed-form approximations have only been obtained for certain pairs of response distributions and link functions. 

In this article, we propose an extended variational approximations (EVA) approach which widens the set of VA-applicable GLLVMs drastically. EVA draws inspiration from the underlying idea of Laplace approximations: by replacing the complete-data likelihood function with its second order Taylor approximation about the mean of the variational distribution, we can obtain a closed-form approximation to the marginal likelihood of the GLLVM for any response type and link function. Through simulation studies and an application to testate amoebae data set in ecology, we demonstrate how EVA results in a universal approach to fitting GLLVMs, which remains competitive in terms of estimation and inferential performance relative to both standard VA and a Laplace approximation approach, while being computationally more scalable than both in practice.
\end{abstract}


\section{Introduction}
\label{sec:intro}
In many scientific disciplines, there is a growing need to process and analyze multi-response or multivariate data. With multiple responses being measured at each observational unit, a key research question and/feature that needs to be accounted for is the underlying structural relationships between the response variables. A prime example of this comes from community ecology, where researchers analyze multivariate abundance data to establish relationships between interacting plant and animal species and the various processes driving these joint distributions \citep{Pollocketal:2014,Ovaskainenetal:2017,wagner2020improved}. Such data can naturally be represented as $n \times m$ matrix $\bm Y$, where element $y_{ij}$ denotes the observation of response $j=1,\dots,m$ recorded at observational unit $i=1,\dots,n$. The types of responses can vary widely, for instance in ecology we may record binary 'presence/absence' responses, overdispersed (and occasionally underdispersed) counts, semi-continuous data e.g., biomass which is non-negative and often has a strong spike at zero, and proportions data between zero and one. With such a variety of response types, it is important that a statistical modeling approach be able to handle these, and account for associated mean-variance relationships \citep{warton2017central}.

Over the past two decades, generalized linear latent variable models \citep[GLLVMs,][]{skrondal2004generalized} have emerged as a powerful class of methods for analyzing multivariate response capable of handling the aforementioned wide variety of response types through an approximate distributional assumption \citep[see for instance,][]{warton2015so,Bjorketal}. In GLLVMs, the mean response is modeled as a function of a set of underlying {latent variable values} or {scores} $\bm{u}_i = (u_{i1}, \dots, u_{ip})^\intercal$, along with any measured predictors as appropriate. By assuming the number of latent variables so that $p << m$, GLLVMs offer a parsimonious way of modeling between-response correlations through rank-reduction. The latent variables themselves may posses some interpretation. For instance, in the field of psychology, these latent variables can be seen as an unobserved measure of traits, that can only be tested for indirectly, such as person's intelligence, anxiety or welfare \citep{moustaki2000generalized}. In community ecology, latent variables are often interpreted as a set ordination axes describing different test sites by their species relative abundance or composition \citep{Huietal:2015,warton2015so,damgaard2020model}.




While it is a powerful approach, in practice fitting GLLVMs remains a computationally burdensome task. Focusing on likelihood-based approaches, the missing latent variables need to be integrated out, and this results in a marginal likelihood function that lacks an tractable solution except in special cases such as normally distributed responses and the identity link function. This computational challenge has spurred a variety of research into approximation schemes to overcome the integral, with popular ones being the Monte-Carlo {Expectation-Maximization algorithm} \citep{wei1990monte}, Laplace approximations and variations thereof including  quadrature methods \citep{Huberetal:2004,bianconcini2012estimation,Nikuetal:2017}, and the more recent {variational approximations} \citep[VA,][]{Huietal:2017,Nikuetal:2019,Zengetal:2021,veen2021} which we focus on in this article. For fitting GLLVMs, VA has been shown to be computationally more efficient and scalable than Laplace approximations and quadrature, and in some situations can be more stable \citep{Nikuetal:2019}. On the other hand, unlike Laplace approximations and other estimation methods such as Bayesian Markov Chain Monte Carlo approaches, the application of VA to GLLVMs and other mixed models so far has been hampered considerably by its lack of the general applicability. 
For example, in the case of Bernoulli responses one is forced to use a probit link function in order to obtain a fully closed-form approximation, otherwise for the canonical logit or other links such as complementary log-log link additional approximations may need to be taken. For some response types such as GLLVMs with the Tweedie distribution, to our knowledge no attempt has even been made to apply VA as little simplifications can be made to facilitate closed-form and thus computationally efficient approximations. 

To address the above issue, in this article we propose an extended variational approximations (EVA) that allows for fast and practical universal fitting of GLLVMs. EVA is inspired by the underlying idea of the Laplace approximation: by replacing the complete-data log-likelihood function with a second-order Taylor series expansion about the mean of the variational distribution, we can obtain a closed-form variational lower bound of the marginal log-likelihood for any response type and link function. We demonstrate how this breakthrough allows VA to be applied many more types of GLLVMs that are commonly applied, say, in community ecology. Furthermore, as with the conventional VA method, EVA allows statistical inference through the adaptation of standard tools such as using the observed information matrix for constructing confidence intervals and hypothesis tests, model selection and residual analysis approaches, and ordination and prediction coupled with associated uncertainty quantification. An extensive simulation study and an application to data set of testate amoebae counts recorded at peatland sites across Finland demonstrates how EVA results in a universal approach to fitting GLLVMs, that is competitive in estimation and inferential performance compared to both standard VA and a Laplace approximation approach while being computationally more scalable than both in practice.

The rest of this article is structured as follows: Section \ref{sec:gllvm} provides an overview of GLLVM and the conventional VA approach. Sections \ref{sec:eva} and \ref{sec:deriv_commontypes} introduce the extended variational approximations (EVA) approach and derives the approximated log-likelihoods using EVA for some several types of GLLVMs respectively. Section \ref{sec:simul} demonstrates the estimation and inferential performance as well as the relative computational efficiency of EVA through a set of three distinct numerical studies, while Section \ref{sec:example} illustrates an application of EVA to a dataset of multivariate abundances of testate amoebae across Finland.
Finally, potential avenues of future research for EVA are discussed in Section \ref{sec:discussion}. 

\section{Generalized linear latent variable models}
\label{sec:gllvm}

Let $\mu_{ij} = \text{E}(y_{ij} | \bm{u}_i)$ denote the conditional expectation for {response} $j = 1, \dots ,m$ at {observational unit} $i =1 \dots, n$, given the vector of latent variables $\bm{u}_i$. We assume that $n$ observational units are independent of each other. The GLLVM is then characterized by the following mean model:
\begin{equation}
\label{eqn:gllvm1}
    g(\mu_{ij}) = \eta_{ij}=\alpha_i + \beta_{0j} + \bm x^\intercal_i\bm\beta_j + \bm u^\intercal_i\bm\lambda_j,
\end{equation}
where $g(\cdot)$ is a known link function, $\bm{x_i}=(x_{i1},\dots,x_{iq})^\intercal$ denotes a $q$-vector of the observed predictors for unit $i$, and $\bm{\beta_j}=(\beta_{j1},\dots,\beta_{jq})^\intercal$ are the corresponding response-specific regression coefficients. As an aside, note that $\mu_{ij}$ is defined conditionally on $\bm{x}_i$ as well as on $\bm{u}_i$, although for ease of notation the former is suppressed. Next, $\beta_{0j}$ denotes the response-specific intercept, while $\alpha_i$ is an optional unit-specific parameter that can be treated either as a fixed or random effect. In the setting of community ecology, \citet{Huietal:2015} and \citet{Bjorketal} among others suggested including $\alpha_i$ in (\ref{eqn:gllvm1}) if an ordination based on species composition is desired, as this effect acts to account for sampling variety and heterogeneity in total abundance across sites.
Finally, the $p$-vector $\bm{\lambda_j}=(\lambda_{j1},\dots,\lambda_{jp})^\intercal$ denotes a set of response-specific loadings which quantify the relationship between the mean response and the latent variables. 

Turning to the latent variables, in (\ref{eqn:gllvm1}), it is most common to assume that the $\bm{u}_i$'s are independent vectors from a standard multivariate normal distribution, $\bm{u}_i \sim N_p(\bm 0, \bm I_p)$, where $\bm I_p$ denotes a $p\times p$ identity matrix. Here the zero mean fixes the locations and the unit variance fixes the scale of latent variables and ensures parameter identifiability \citep[Chapter 5,][]{skrondal2004generalized}. Furthermore, if we consider the $m \times p$ loading matrix formed by stacking the $\bm{\lambda}_j$'s as vectors, $\bm\Gamma = [\bm{\lambda_1} \dots \bm{\lambda_m}]^\intercal$, then to ensure that the parameters are identifiable it is common to also constrain the upper triangular component of $\bm\Gamma$ to zero and the diagonal elements to be positive \citep{Huberetal:2004,Nikuetal:2017}. Note that such constraints do not restrict the flexibility of the GLLVM: specifically, the latent variable component $\bm u^\intercal_i \bm{\lambda_j}$ in equation (\ref{eqn:gllvm1}) accounts for any residual correlation not accounted for by the covariates $\bm{x}_i$, such that the residual $m \times m$ covariance matrix on the linear predictor scale is given by $\bm\Sigma = \bm\Gamma\bm\Gamma^\intercal$. We thus see that GLLVM models the covariance between the responses via rank-reduction, and the choice of the number of latent variables $p$ can vary depending on the aim of the GLLVM, e.g., \citet{Huietal:2015} and \citet{warton2015so} used $p = 1,2,3$ for the purposes of ordination, while \cite{tobler2019joint} suggested larger values if the goal is to make inference on the $\bm{\beta}_j$'s while accounting for residual correlation between species; see also \citet{hui2018order} for an example of a data-driven approach to choosing $p$.

To complete the formulation of GLLVMs, we assume that the responses $(y_{i1}, \dots, y_{im})^\intercal$ are conditionally independent given the vector of latent variables $\bm{u}_i$. 
Specifically, let $\bm \Psi = (\bm \alpha^\intercal, \bm \phi^\intercal, \bm \beta_0^\intercal,\bm \beta_1^\intercal, \ldots, \bm \beta_m^\intercal, \vect(\bm \Gamma)^\intercal)^\intercal$ denote the vector of all model parameters in the GLLVM, where $\bm \alpha = (\alpha_1, \dots, \alpha_n)^\intercal$, $\bm \beta_0 = (\beta_{01}, \dots, \beta_{0m})^\intercal$, and $\bm \phi = (\phi_1, \ldots, \phi_m)^\intercal$ denote a vector of nuisance parameters used also to characterize the conditional distribution of the responses. These may be known \emph{a-priori} or may also need to estimated. Let $\bm u = (\bm u_1^\intercal, \dots, \bm u_n^\intercal)^\intercal$ denote the full $np$-vector of the latent variables. Then the complete-data likelihood function for a GLLVM is defined as
\begin{align}
\label{eqn:completedatalikelihood}
    L(\bm \Psi; \bm u) &= \prod_{i=1}^{n}\left(\prod_{j=1}^{m}f(y_{ij}|\bm{u}_i, \bm \Psi)\right)f(\bm{u}_i) = f(\bm y|\bm u, \bm \Psi)f(\bm u),
\end{align}
where $f(y_{ij}| \bm{u}_i, \bm \Psi)$ denotes the conditional distribution of $y_{ij}$ and $f(\bm{u}_i) = N_p(\bm 0, \bm I_p)$. As discussed previously, one of the main strengths of GLLVM is the capacity to handle a wide variety of response types (and associated mean-variance relationships), and this is done by selecting an appropriate form for $f(y_{ij}| \bm{u}_i, \bm \Psi)$; see Section \ref{sec:deriv_commontypes} for some examples of particular relevance in ecology. Also, note that while we have assumed that all the $y_{ij}$'s follow the same distributional form, this need not be the case and the developments of EVA below can be straightforwardly extended to the case where the $m$ responses are of different types \citep[e.g.,][]{Sammeletal:1997}.

Based on (\ref{eqn:completedatalikelihood}), we then obtain the marginal likelihood function by integrating over the random latent variables $L(\bm \Psi) = \int f(\bm y|\bm u, \bm \Psi)f(\bm u)d(\bm u)$. Maximum likelihood estimates are then calculated as $\arg\max_{\bm \Psi} L(\bm \Psi)$. Optimizing the marginal likelihood function however presents a major computational challenge, as the integral does not possess a closed form except for special cases such when the $y_{ij}$'s are (also) normally distributed and $g(\cdot)$ is set to the identity link. To overcome this, and focusing on likelihood-based methods, a variety of approximation approaches have been proposed as reviewed in Section \ref{sec:intro}. A more recent approach, which we focus on in this article, is that of variational approximations. 


\subsection{Variational approximations}
\label{sec:va}
{Variational approximations} (VA) refers to a general class of methods that originated in the machine learning literature, and were subsequently popularized in statistics by \cite{ormerod10} and \citet{blei2017variational} among others. Most of the research into VA has been that of variational Bayes, although in this article we will be focusing on likelihood-based estimation and approximations of the marginal log-likelihood function. VA was proposed for likelihood-based estimation of generalized linear mixed models initially by \cite{ormerod12}, and has since been studied and applied in various mixed models settings \citep{SiewNott:2014,LeeWand:2016,Nolanetal:2020} and semiparametric regression including generalized additive models \citep{Lutsetal:2014,MenictasWand:2015,Huietal:2018}. For GLLVM specifically, VA was first proposed by \citet{Huietal:2017}, and has since been further studied by \citet{Nikuetal:2019}, \citet{veen2021}, and \citet{Zengetal:2021}. Note however that the many of these developments have focused on a limited number of response types and link functions. 

The basic idea of variational approximation is to develop a so-called {variational lower bound} to the marginal log-likelihood function, also known as the VA log-likelihood function. In the context of GLLVMs, this is developed as
\begin{align} \label{eqn:va1}
\log L(\bm \Psi) &= \log \left(\int f(\bm y|\bm u, \bm \Psi)f(\bm u)d(\bm u)\right) \geq  \int q(\bm u)\log \left\{ \frac{f(\bm y|\bm u, \bm \Psi)f(\bm u)}{q(\bm u)} \right\}d(\bm u) \triangleq \underline{\ell}(\bm \Psi | q),
\end{align} 
where $q(\bm u)$ denotes the density of the assumed variational distribution of the latent variables $\bm u$. For a more detailed derivation and explanation of (\ref{eqn:va1}), we refer readers to \cite{ormerod10} and references therein. Note that equality is achieved if we set $q(\bm u) = f(\bm{u}|\bm{y},\bm{\Psi})$, i.e., the posterior distribution of the latent variables. However in general like in the marginal likelihood function this does not posses a tractable form, and so in VA we further assume that the variational distribution belongs to some parametric family of distributions $\{q(\bm u | \bm \xi) : \bm \xi \in \bm \Xi\}$ for a set of variational parameters $\bm{\xi}$. For GLLVMs in particular, and given the form of the complete-data likelihood function in (\ref{eqn:completedatalikelihood}), \citet{Huietal:2017} among others employed a mean-field assumption and set $q(\bm u | \bm{\xi}) = \prod_{i=1}^n q_i(\bm{u}_i | \bm{\xi}_i)$ where $q_i(\bm{u}_i | \bm{\xi}_i) = N_p(\bm{a}_i, \bm{A}_i)$. That is, the variational distribution of the latent variables for unit $i$ is assumed to be multivariate normal distribution with mean vector $\bm{a}_i$ and covariance matrix $\bm{A}_i$. \citet{Huietal:2017} in fact showed that this choice of $q(\cdot)$ was optimal (in a Kullback-Leibler divergence sense) among the family of multivariate normal distributions. Applying this form of the variational distribution to (\ref{eqn:va1}), we thus obtain 
{\normalsize
\begin{align}
\label{eqn:va2}
    \underline{\ell}(\bm \Psi, \bm \xi | q) &= \int q(\bm u | \bm \xi)\log f(\bm y | \bm u, \bm \Psi) d(\bm u) + \int q(\bm u | \bm \xi)\log\left\{\frac{f(\bm u)}{q(\bm u | \bm \xi)}\right\} d(\bm u) \nonumber \\
    &= \int q(\bm u | \bm \xi)\log f(\bm y | \bm u, \bm \Psi) d(\bm u) + \frac{1}{2} \sum_{i=1}^n \big\{ \log \det (\bm{A}_i)  - \bm{a}_i^\intercal \bm{a}_i - \Tr(\bm{A}_i)\big\},
\end{align}
}%
where the last line follows from well-known formulas relating to the entropy of a multivariate normal distribution, and constants with respect to the model and variational parameters are omitted. By treating (\ref{eqn:va2}) as the new objective function and then solving $(\widehat{\bm \Psi}, \widehat{\bm \xi}) = \underset{\bm \Psi, \bm \xi}{\argmax} \: \underline{\ell}(\bm \Psi, \bm \xi | q)$, we then obtain VA estimates of the model parameters $\widehat{\bm \Psi}$ as well as those of the variational parameters $\hat{\bm{\xi}}$. Indeed, once the GLLVM is fitted using VA, the estimated variational distributional distributions $\hat{q}_i(\bm{u}_i) = N_p(\hat{\bm{a}}_i, \hat{\bm{A}}_i)$ is then an approximation of the true posterior distribution of the latent variables. 

As an approach, VA (and variational Bayes) has previously been shown in many contexts to provide a strong balance between estimation accuracy and computational efficiency/scalability; see \citet{ormerod10} and \citet{Nikuetal:2019} among many others for a variety of simulations, as well as the asymptotic theory of \citet{hall2011theory} and \citet{wang2019frequentist} among others. On the other hand, to facilitate this computational efficiency, we ideally require a closed-form expression for the first term on the right hand side of \eqref{eqn:va2}. Unfortunately, in general this is not guaranteed, and so the development of VA for GLLVMs has so far been limited to selected response distributions or link functions, limiting the applicability of the approach.

\section{Extended variational approximations} \label{sec:eva}
As reviewed above, one drawback of the standard VA approach for GLLVMs is that the exact formulation of the variational lower bound in (\ref{eqn:va2}) depends heavily on the assumed distribution for the responses $f(y_{ij}, \bm{u}_i, \bm{\Psi})$ and the associated link function $g(\cdot)$. A fully tractable form is not always available, even with some of the more popular response-link combinations. A prime example is the case of Bernoulli distributed responses, where the probit link function is known to lead to fully closed-form variational lower bound, but the canonical logit link or the complementary log-log link do not \citep[and has led to various additional approximations being made, e.g.,][]{blei2007correlated,Huietal:2018}. Another example is GLLVM with Tweedie distributed responses, where to our knowledge VA does not lead to a fully-closed form approximation. In turn, this means alternate approaches such as Laplace approximations has to be used instead for such models \citep{Nikuetal:2017}.


To overcome the above issues and further broaden the applicability of VA as computationally efficient approach to fitting GLLVMs, we propose an approach called {extended variational approximation} or EVA. The method is similar that of delta method variational inference as proposed by \citet{wang2013variational}, although to our knowledge this article is first to apply it for GLLVMs. The idea of EVA is similar to and indeed inspired by that of the Laplace approximation. Specifically, we replace the complete-data log-likelihood function $\log L(\bm{\Psi};\bm{u})$ by its a second-order Taylor expansion with respect to the latent variables $\bm u$. In the case of GLLVMs, because the latent variables are assumed to be normally distributed then we need only perform the expansion on the log-density of the responses $\log f(\bm y | \bm u, \bm \Psi)$. Importantly, in EVA this expansion is taken around the mean of the variational distribution i.e., $\bm{a} = (\bm a_1^\intercal, \ldots, \bm a_n^\intercal)^\intercal$, which serves as a natural center point of the approximation. 
{\normalsize
\begin{align*}
    &\log f(\bm y|\bm u, \bm \Psi) \approx \log f(\bm y|\bm a, \bm \Psi) + (\bm u - \bm a)^\intercal \, \frac{\partial \log f(\bm y|\bm u, \bm \Psi)}{\partial \bm{u}}\bigg|_{\bm u = \bm a} + \frac{1}{2}(\bm u - \bm a)^\intercal \bm{H}(\bm{a},\bm{\Psi}) (\bm u - \bm a),
\end{align*}
where $\bm{H}(\bm{a},\bm{\Psi}) = \partial^2 \log f(\bm y|\bm u, \bm \Psi)/\partial \bm{u} \partial \bm{u}^\intercal \big|_{\bm u = \bm a}$. 
}%
By substituting the above expansion into (\ref{eqn:va2}) and noting that $\int (\bm u - \bm a)^\intercal \, \partial \log f(\bm y|\bm u, \bm \Psi)/\partial \bm{u} \big|_{\bm u = \bm a} q(\bm u | \bm \xi) \, d \bm{u} = \bm{0}$, we thus obtain the extended variational lower bound, which we also refer to as the EVA log-likelihood for GLLVMs,
\begin{align}
\label{eqn:eva3}
    \underline{\ell}(\bm \Psi, \bm \xi) &\approx \int q(\bm u | \bm \xi)\Big\{ \log f(\bm y|\bm a, \bm \Psi) + (\bm u - \bm a)^\intercal \, \frac{\partial \log f(\bm y|\bm u, \bm \Psi)}{\partial \bm{u}}\bigg|_{\bm u = \bm a} \nonumber \\  
    &\phantom{=} +\frac{1}{2}(\bm u - \bm a)^\intercal\,\bm{H}(\bm{a},\bm{\Psi})\,(\bm u - \bm a) \Big\}d(\bm u) \nonumber \\ 
    &\phantom{=}+ \frac{1}{2} \sum_{i=1}^n \big\{ \log \det (\bm{A}_i)  - \bm{a}_i^\intercal \bm{a}_i - \Tr(\bm{A}_i)\big\} \nonumber \\
    &= \sum_{i=1}^n \sum_{j=1}^m \log f(y_{ij} | \bm{a}_i, \bm \Psi) + \frac{1}{2}\Tr(\bm{H}_i(\bm{a}_i,\bm{\Psi}) \bm{A}_i) \nonumber \\
    &\phantom{=}+ \frac{1}{2} \sum_{i=1}^n \big\{ \log \det (\bm{A}_i)  - \bm{a}_i^\intercal \bm{a}_i - \Tr(\bm{A}_i)\big\} \triangleq \ell_{\mathrm{EVA}}(\bm \Psi, \bm \xi),
\end{align}
where $$\bm{H}_i(\bm{a}_i,\bm{\Psi}) = \partial^2 \sum_{j=1}^m \log f(y_{ij} | \bm{u}_i, \bm \Psi)/\partial \bm{u}_i \partial \bm{u}_i^\intercal \big|_{\bm{u}_i = \bm{a}_i}.$$ Likelihood-based EVA estimates for both model and variational parameters are then obtained by maximizing \eqref{eqn:eva3}. That is, $(\hat{\bm \Psi}_\mathrm{EVA}, \hat{\bm \xi}_\mathrm{EVA}) = \underset{\bm \Psi, \bm \xi}{\argmax} \: \ell_\mathrm{EVA}(\bm \Psi, \bm \xi)$. Importantly, as can be seen there are no integrals in $\ell_{\mathrm{EVA}}(\bm \Psi, \bm \xi)$, meaning its maximization can be done using generic optimization approaches, say. This allows EVA to be applied universally to all response types and link functions of GLLVMs, assuming an appropriate parametric form for the conditional distribution of the former, and in Section \ref{sec:deriv_commontypes} we will provide some examples of applying EVA to common types of GLLVMs seen in community ecology among other disciplines. At the same time, it inherits the computational efficiency and scalability of the standard VA approach, as will be seen in the simulation studies in Section \ref{sec:simul}.


\subsection{Inference and ordination}
\label{sec:inference}
Similar to the standard VA method \citep[e.g.,][]{Huietal:2017}, we can adapt many of the existing likelihood-based approaches to statistical inference for EVA. In section, we briefly discuss some of these. First, after fitting we can calculate approximate standard errors for the estimates of the model parameters in the GLLVM based on the observed information matrix. That is, we first calculate 
\begin{align*}
\bm{I}(\hat{\bm \Psi}_\mathrm{EVA}, \hat{\bm \xi}_\mathrm{EVA}) = -\frac{\partial^2 \ell_{\mathrm{EVA}}(\bm{\Psi},\bm{\xi})}{\partial (\bm{\Psi},\bm{\xi}) \partial (\bm{\Psi},\bm{\xi})^T}\Bigg|_{\bm{\Psi}=\hat{\bm{\Psi}}_\mathrm{EVA},\,\bm{\xi}=\hat{\bm{\xi}}_\mathrm{EVA}}.
\end{align*}
Then the relevant block of matrix $\bm{I}(\hat{\bm \Psi}_\mathrm{EVA}, \hat{\bm \xi}_\mathrm{EVA})^{-1}$, corresponding to the model parameters, denoted here as $\bm{I}(\hat{\bm \Psi}_\mathrm{EVA}, \hat{\bm \xi}_\mathrm{EVA})^{-1}_{\bm{\Psi}}$, provides approximate standard errors for $\hat{\bm{\Psi}}_\mathrm{EVA}$. Approximate Wald-based confidence intervals as well as conduct Wald tests for the model parameters can be then constructed as usual. Alternatively, likelihood ratio tests and corresponding confidence intervals can also be developed based on the (maximized value of the) EVA log-likelihood $\ell_{\mathrm{EVA}}(\bm \Psi, \bm \xi)$.

Another popular application of GLLVMs, particularly in ecology, is that of ordination. Using EVA, we can use the estimated mean vectors of the variational distribution, $\hat{\bm a}_i$, $i=1,\dots,n$, as point predictions of the latent variables $\bo u_i$, which can then be plotted as a means of model-based unconstrained or residual ordination. Similar to \citet{ormerod12} and \citet{Huietal:2017}, $\hat{\bm a}_i$ from EVA can be regarded as variational version of the empirical Bayes predictor and maximum \emph{a-posteriori} predictor (MAP) of the latent variable. A biplot can also be constructed by including the estimated loadings $\hat{\bm \lambda}_j$, $j=1,\ldots,m$, on the same ordination. Next, regarding uncertainty quantification, note that the estimated covariance matrices $\hat{\bm A}_i$, $i=1,\ldots,n$, provide an estimate of the posterior covariance of the latent variables, and can be thus used to obtain prediction regions for the latent variables. However, in practice these tend to underestimate the true posterior covariance as they fail to account for the uncertainty of the estimated parameters. To overcome this, we can develop a variational analogue of the solution developed by \cite{BoothHobert:1998} based on conditional mean squared errors of predictions (CMSEP). Specifically, we can approximate the CMSEP as
\begin{align*}
CMSEP(\bo{\hat a}_i;\bo{\Psi},\bo y_i)  &= \mathbb{E}\left\{(\hat{ \bm a}_i - \bm u_i)(\hat{ \bm a}_i - \bm u_i)^\intercal|\bm{y}_i \right\} \approx \bo{\hat A}_i + \bm{\hat Q} \, \bm{I}(\hat{\bm \Psi}_\mathrm{EVA}, \hat{\bm \xi}_\mathrm{EVA})^{-1}_{\bm{\Psi}}\, \bm{\hat Q}^\intercal,
\end{align*}
where $\bm{\hat Q} = \bm{Q}(\bm{\hat \Psi}_\mathrm{EVA}, \bm{\hat \xi}_\mathrm{EVA})$, with $$\bm{Q}(\bm \Psi, \bm \xi) = \left(\frac{\partial^2  \ell_{\mathrm{EVA}}(\bo{ \Psi},\bo{\xi})}{ \partial \bo a_i\partial \bo a_i^\intercal}\right)^{-1} \left(\frac{\partial^2  \ell_{\mathrm{EVA}}(\bo{ \Psi},\bo{\xi})}{\partial \bo a_i \partial \bo\Psi^\intercal} \right).$$ Prediction regions can then be constructed for the latent variables based on using $CMSEP(\bo{\hat a}_i;\bo{\Psi},\bo y_i)$ as the approximate standard error. 

Using the EVA estimates, we can perform residual analysis to assess whether there are major violations in the assumptions underlying the GLLVM, in much the same way as with other common regression models. For instance we can calculate Dunn-Smyth residuals \citep{dunn1996randomized} to construct residual diagnostic plots such as residual versus fitted values and normal quantile-quantile plots, where these residuals are defined as
$r_{q,ij} = \bm \Phi^{-1}(c_{ij})$ and $$c_{ij} \sim \text{Unif}\left(\lim_{y \uparrow y_{ij}} F(y | \hat{\bm{u}}_{i}, \hat{\bm{\Psi}}), F(y_{ij}| \hat{\bm{u}}_{i}, \hat{\bm{\Psi}}\right),$$ with $F(\cdot|\bm{u}_{i},\bm{\Psi})$ denoting the cumulative distribution function of the response $y_{ij}$. If the underlying assumptions of the GLLVM are reasonably well satisfied, then the Dunn-Smyth residuals should follow a standard normal distribution.

The above describes only some of various statistical inferences that a practitioner may wish to drawn from an GLLVM. There are many others possible, e.g., model selection using information criteria or regularization methods, but more importantly we emphasize that all of these tools are adaptable to the setting where EVA is used to fit GLLVMs. Indeed, by adapting such tools over and studying their properties as avenues of future research, it further strengthens the the universality of EVA as an approach to estimation and inference for GLLVMs.


We conclude this section with a short note regarding implementation. We implemented EVA using a combination of $\mathsf{R}$ and in C\texttt{++} via the package $\mathsf{TMB}$ \citep{tmb}. That is, the (negative) EVA log-likelihood function for the relevant GLLVM were first written in C\texttt{++}, after which there are compiled by $\mathsf{TMB}$ which employs automatic differentiation to produce $\mathsf{R}$ functions to calculate the negative log-likelihood, the score, and potentially the Hessian matrix. We then pass these to a generic optimization procedure such as the function $\mathsf{optim}$ to maximize the EVA log-likelihood and calculate the observed information matrix. The CMSEP can be calculated in a similar manner. The full implementation of EVA is available as part of the 
package $\mathsf{gllvm}$ \citep{Nikuetal:2019b}. As starting values for EVA, we use the proposal in \citet{Nikuetal:2019}.

\section{EVA for some common types of GLLVMs} \label{sec:deriv_commontypes}

In this section, we present some forms of the EVA log-likelihood for combinations of response distributions and link functions that are commonly used with GLLVMs, especially in the context of community ecology. 
We begin by formulating the EVA log-likelihoods when the responses $y_{ij}$ are assumed to come from the one-parameter exponential family of distributions. All proofs are provided in Appendix \ref{appendix:proofs}.

\begin{restatable}{theorem}{thmone}
\label{thm:thm1}
For the GLLVM with mean model given by (\ref{eqn:gllvm1}), let the conditional distribution of the responses be part of the exponential family, $f(y_{ij}|\bm{u}_i, \bm \Psi) = \exp\left\{h_j(y_{ij})b_j(\mu_{ij}) - c_j(\mu_{ij}) + d_j(y_{ij}) \right\}$ for known functions $h_j(\cdot)$, $b_j(\cdot), c_j(\cdot)$ and $d_j(\cdot)$. If $b_j(\cdot)$ and $c_j(\cdot)$ as well as the link function $g(\cdot)$ are at least twice continuously differentiable with $g'(\mu_{ij})\neq 0$, then the EVA log-likelihood in \eqref{eqn:eva3} takes the closed-form
{
\begin{align*}
    \ell_{\mathrm{EVA}}(\bm \Psi, \bm \xi) &= \sum_{i=1}^n \sum_{j=1}^m \log f(y_{ij} | \bm{a}_i, \bm \Psi) 
    + \frac{1}{2} \sum_{i=1}^n \Big\{ \log \det (\bm{A}_i)  - \bm{a}_i^\intercal \bm{a}_i - \Tr(\bm{A}_i)\Big\} \nonumber \\ 
    &\phantom{=} + \frac{1}{2}\sum_{i=1}^n \sum_{j=1}^m \left[\left\{\frac{h_j(y_{ij})b_j''(\tilde\mu_{ij})-c_j''(\tilde\mu_{ij})}{(g'(\tilde\mu_{ij}))^2} - \frac{h_j(y_{ij})b_j'(\tilde\mu_{ij})-c_j'(\tilde\mu_{ij})}{\big(g'(\tilde\mu_{ij}) \big)^3/g''(\tilde\mu_{ij})}  \right\}\bm\lambda_j^\intercal \bm{A}_i \bm \lambda_j \right],
\end{align*}
}%
where $\tilde\mu_{ij} = g^{-1}(\tilde\eta_{ij}) = g^{-1}( \alpha_i + \beta_{0j} + \bm x^\intercal_i\bm\beta_j + \bm a^\intercal_i\bm\lambda_j)$.
\end{restatable}

Some simplifications can be made in the case where the canonical link function is used. 
\begin{restatable}{corollary}{corone}
\label{thm:cor1}
If in Theorem \ref{thm:thm1} the link function is taken to be the canonical link function, i.e., $g \equiv b$, then the EVA log-likelihood reduces to
{
\begin{align*}
    \ell_{\mathrm{EVA}}(\bm \Psi, \bm \xi) &= \sum_{i=1}^n \sum_{j=1}^m \log f(y_{ij} | \bm{a}_i, \bm \Psi) 
   + \frac{1}{2} \sum_{i=1}^n \Big\{ \log \det (\bm{A}_i)  - \bm{a}_i^\intercal \bm{a}_i - \Tr(\bm{A}_i)\Big\} \nonumber \\ 
    &\phantom{=} + \frac{1}{2}\sum_{i=1}^n \sum_{j=1}^m \left[\left\{\frac{b_j''(\tilde\mu_{ij})c_j'(\tilde\mu_{ij}) - b_j'(\tilde\mu_{ij})c_j''(\tilde\mu_{ij})}{(b_j'(\tilde\mu_{ij}))^3}  \right\}\bm\lambda_j^\intercal \bm{A}_i \bm \lambda_j \right].
\end{align*}
}%
\end{restatable}
From Theorem \ref{thm:thm1}, we see that EVA shares the same term $\log \det (\bm{A}_i)  - \bm{a}_i^\intercal \bm{a}_i - \Tr(\bm{A}_i)$ as the standard VA log-likelihood for GLLVMs \citep{Huietal:2017, Nikuetal:2019}, but also involves computation of a Hessian term based on the conditional distribution of the response \citep{Huberetal:2004}. 

\subsection{Overdispersed counts}
\label{subsec:nbgllvm}
Multivariate count data are one of the most common applications of GLLVMs, with a starting choice often being to assume that counts follow a Poisson distribution with the canonical log link. However in many settings such as community ecology and microbiome applications, the Poisson model is often inappropriate due to the prevalence of overdispersion. Therefore, a popular alternative for overdispersed counts is to consider negative binomial GLLVMs with the log link, where $$f(y_{ij}|\bm{u}_i, \bm \Psi) = \frac{\Gamma(y_{ij}+\phi_j^{-1})}{\Gamma(\phi_j^{-1})y_{ij}!} \frac{\left\{\phi_j\mu_{ij}\right\}^{y_{ij}}}{\left\{\phi_j\mu_{ij}+1\right\}^{y_{ij}+\phi_j^{-1}}} $$ and $\phi_j > 0$ is the response-specific overdispersion parameter. The mean-variance relationship is quadratic in form, $\Var(y_{ij}) = \mu_{ij} + \phi_j\mu_{ij}^2$, making it suitable for handling a considerable amount of overdispersion. 

Previously, \citet{Huietal:2017} considered negative binomial GLLVMs using the standard VA approach, but had to reparameterize the negative binomial distribution as a Poisson-Gamma mixture in order to derive closed-form variational lower bound \citep[see also][]{Zengetal:2021}. With EVA however, this is not necessary and we can explicitly use the form presented above as follows. 
If we assume assume $\phi_j$ is fixed, then the negative binomial distribution is part of the exponential family. Therefore we can apply Theorem \ref{thm:thm1} to straightforwardly show that EVA log-likelihood takes the following form for negative binomial GLLVMs:
\begin{align*}
    &\ell_{\mathrm{EVA}}(\bm \Psi, \bm \xi) \\
    &= \sum_{i=1}^n\sum_{j=1}^m \bigg\{ \log \Gamma\left(y_{ij}+\frac{1}{\phi_j}\right) - \log \Gamma\left(\frac{1}{\phi_j}\right) - \frac{1}{\phi_j}\log(\phi_j) + y_{ij}\tilde\eta_{ij} - \left(y_{ij}+\frac{1}{\phi_j}\right)\log\left(\tilde\mu_{ij} + \frac{1}{\phi_j}\right) \bigg\} \\
    &\phantom{=}- \sum_{i=1}^n\sum_{j=1}^m \left\{\frac{\tilde{\mu_{ij}}(1+\phi_j y_{ij})}{2(1+\phi_j\tilde\mu_{ij})^2} \bm \lambda_j^\intercal \bm{A}_i \bm \lambda_j \right\} + \frac{1}{2} \sum_{i=1}^n \Big\{ \log \det (\bm{A}_i)  - \bm{a}_i^\intercal \bm{a}_i - \Tr(\bm{A}_i)\Big\},
\end{align*}
where $\tilde \mu_{ij}= \exp(\tilde \eta_{ij}) = \exp(\alpha_i + \beta_{0j} + \bm x^\intercal_i\bm\beta_j + \bm a^\intercal_i\bm\lambda_j)$.
The above result can be applied in a more general setting where the $\phi_j$'s also need to be estimated, by deriving the Hessian $\bm{H}_i(\bm{a}_i,\bm{\Psi})$ directly.

\subsection{Binary responses}
\label{subsec:binary}
For binary responses, e.g., presence-absence data in community ecology, we can assume that $y_{ij}$ follows a Bernoulli distribution, i.e. $f(y_{ij}|\bm{u}_i, \bm \Psi) = \mu_{ij}^{y_{ij}} \{1-\mu_{ij}\}^{1-y_{ij}}$. 
The Bernoulli distribution belongs to the exponential family meaning we can apply Theorem \ref{thm:thm1} to obtain the EVA log-likelihood irrespective of the link function assumed. Below, we discuss two of the most commonly used links used.


Bernoulli GLLVMs with the canonical logit link, $\mathrm{logit}(\mu_{ij}) = \log\{\mu_{ij}/(1-\mu_{ij})\} = \eta_{ij}$, presents a good example of a situation where the standard VA approach, as described in Section \ref{sec:va}, fails to provide a fully closed-form approximation of the log-likelihood function. Instead, further approximations are required in order to produce a tractable form \citep{blei2007correlated,Huietal:2018}. By contrast, for EVA we can directly apply Corollary \ref{thm:cor1} and obtain the following EVA log-likelihood function for binary GLLVMs with the logit link:
    \begin{align*}
        \ell_{EVA}(\bm\Psi, \bm\xi) &= \sum_{i=1}^n\sum_{j=1}^m \Bigg\{y_{ij} \tilde\eta_{ij} - \log\{1+\exp({\tilde\eta_{ij})}\}
        - \frac{\exp(\tilde\eta_{ij})}{2\{1+\exp(\tilde\eta_{ij})\}^2} \, \bm \lambda_j^\intercal \bm{A}_i \bm \lambda_j \Bigg\} \nonumber \\
                             &\quad + \frac{1}{2} \sum_{i=1}^n \Big\{ \log \det (\bm{A}_i)  - \bm{a}_i^\intercal \bm{a}_i - \Tr(\bm{A}_i)\Big\},
    \end{align*}
where $\tilde\eta_{ij} = \alpha_i + \beta_{0j} + \bm x^\intercal_i\bm\beta_j + \bm a^\intercal_i\bm\lambda_j$.

To circumvent the above issues with using logit link, \citet{ormerod10} and \citet{Huietal:2017} among others instead considered binary GLLVMs using the probit link $\Phi^{-1}(\mu_{ij})  = \eta_{ij}$, where $\Phi$ is the cumulative distribution function of the standard normal distribution. With this choice of link function standard VA can yield a fully-closed form approximation based on augmenting the complete-data likelihood function with an additional intermediate standard normal random variable and employing the dichotomization trick. By contrast,
with EVA binary GLLVMs using the probit link function follow directly from applying Theorem \ref{thm:thm1}. This leads to the following EVA log-likelihood function:
\begin{align*}
    \ell_{EVA}(\bm\Psi, \bm\xi) &= \sum_{i=1}^n\sum_{j=1}^m \Bigg(y_{ij} \log\big(\tilde\mu_{ij}\big) + (1-y_{ij})\log\big(1-\tilde\mu_{ij}\big) \nonumber \\[0.5ex]
    &\phantom{==} + \frac{1}{2}\Bigg[\phi(\tilde\eta_{ij})^2 \left(\frac{2\tilde\mu_{ij}y_{ij}-y_{ij}-\tilde\mu_{ij}^2}{\tilde\mu_{ij}^2(1-\tilde\mu_{ij})^2}\right) + \phi'(\tilde\eta_{ij})\left\{\frac{y_{ij}-\tilde\mu_{ij}}{\tilde\mu_{ij}(1-\tilde\mu_{ij})}\right\}\Bigg] \bm \lambda_j^\intercal \bm{A}_i \bm \lambda_j\Bigg)
    \nonumber \\[0.5ex]
    &\phantom{=} + \frac{1}{2} \sum_{i=1}^n \Big\{ \log \det (\bm{A}_i)  - \bm{a}_i^\intercal \bm{a}_i - \Tr(\bm{A}_i)\Big\}.
\end{align*}
where $\phi(\tilde\eta_{ij})$ and $\phi'(\tilde\eta_{ij})$ denote the density function of the standard normal distribution and its first derivative, respectively, evaluated at $\tilde\eta_{ij} = \alpha_i + \beta_{0j} + \bm x^\intercal_i\bm\beta_j + \bm a^\intercal_i\bm\lambda_j$.


The above results extend straightforwardly to the case of binomial responses with more than one trial. Also, using Theorem \ref{thm:thm1} EVA can be easily adapted to other link functions such as the complementary log-log link. 

\subsection{Semi-continuous data}
\label{subsec:tweedie}

One type of semi-continuous responses frequently encountered in community ecology, and hence modeled using GLLVMs, is biomass. Derived from recording the total weight of a species at a site, biomass data are non-negative and continuous by nature, often with a large spike at zero (as many species may only be detected at a small number of sites). To model such responses, the Tweedie distribution is often used \citep{Foster2013apm,Nikuetal:2017}, which, for a \textit{power parameter} $1<\nu<2$, can also be parameterized as a compound Poisson-Gamma distribution. Its log-density can be written piecewise as follows: 
\begin{align*}
    \log f(y_{ij}|{\bo u_i}, \bo \Psi) = 
        \begin{cases}
        -\frac{\mu_{ij}^{2-\nu}}{\phi_j(2-\nu)}, & y_{ij} = 0 \\
        \log W(y_{ij}, \phi_j, \nu) + \frac{1}{\phi_j} \left(\frac{y_{ij}\mu_{ij}^{1-\nu}}{1-\nu} - \frac{\mu_{ij}^{2-\nu}}{2-\nu}\right) - \log(y_{ij}) & y_{ij} > 0,
    \end{cases}
\end{align*}  
where $W(y_{ij}, \phi_j, \nu)$ is a generalized Bessel function whose evaluation involves an infinite sum and needs to be evaluated numerically, for example using the method described in \citet{dunn2005}. The Tweedie distribution admits a power-form mean-variance relationship that is generally regarded to be appropriate for biomass data, i.e., $\Var(y_{ij}|\bm{u}_i, \bm{\Psi}) = \phi_j \mu_{ij}^{\nu}$, where $\phi_j > 0$ is a response-specific dispersion parameter and noting that power parameter is usually assumed to be common across responses \citep{Foster2013apm}. A log link function is commonly used with the Tweedie distribution.

To our knowledge, applying standard VA to Tweedie GLLVMs produces no closed-form approximation to the marginal log-likelihood, and as a result limited implementation has taken place, with practitioners instead relying on other approaches such as the Laplace approximation. By contrast, it is straightforward to apply EVA to the Tweedie GLLVMs with the log link: the second order derivative of $\log f(y_{ij}|\bm{u}_i, \bm \Psi)$ with respect to $\bm{u}_i$ are straightforward to calculate (see Appendix \ref{appendix:proofs} for details), after which we can substitute these into \eqref{eqn:eva3} to produce a fully closed-form EVA log-likelihood function.

\subsection{Proportions data}
\label{subsec:beta}


Proportion or percentage data are defined as continuous responses lying in the open unit interval $(0,1)$. In the context of ecology, these responses usually represent the percent cover of plant species at a site \citep{damgaard2019using}. Another example comes from social statistics, say, where we may consider the proportion of household income that is spent on food \citep{ferrari2004beta}. 
If the responses can not take values exactly equal to zero and one, then a common choice to model such multivariate proportions data is to use a beta GLLVM. Specifically, the log-density of the beta distribution is written as
\begin{align*}
    \log f(y_{ij}|{\bm{u}_i}, \bm \Psi) &= \log\Gamma(\phi_j) - \log\Gamma(\mu_{ij}\phi_j) - \log\Gamma\big\{(1-\mu_{ij})\phi_j\big\} \\ 
    &\phantom{=}+ (\mu_{ij}\phi_j-1)\log(y_{ij}) + \big\{(1-\mu_{ij})\phi_j-1\big\}\log(1-y_{ij}),
\end{align*}
where the $\phi_j > 0$ is a response-specific dispersion parameter, and the corresponding mean-variance relationship is given by $\Var(y_{ij}|\bm{u}_i, \bm{\Psi}) = \mu_{ij}(1-\mu_{ij})/(1+\phi_j)$.

As is the case with the Tweedie distribution, applying standard VA to a beta GLLVM fails to admit a closed-form approximation to the marginal log-likelihood, thus hampering its usage. On the other hand, EVA can be easily applied: if we assume say, the logit link function, then we can again 
calculate the second order derivative of $\log f(y_{ij}|\bm{u}_i, \bm \Psi)$ with respect to $\bm{u}_i$ (see Appendix \ref{appendix:proofs} for details). Afterwards, we can substitute these into \eqref{eqn:eva3} to produce a fully closed-form EVA log-likelihood function.

We conclude this section by pointing out that the above only covers a small set of response distributions and link functions, motivated strongly by applications in community ecology, that may use as part of fitting GLLVMs to multivariate data. More importantly, it serves to demonstrate at EVA allows for a more universal approach to variational estimation and inference for GLLVMs, and future computational research will look to expand the implementation EVA even more, e.g., zero-inflated and hurdle type distributions, and cases where the $m$ responses are of different types.

\section{Simulation studies}
\label{sec:simul}

We conducted an extensive simulation study to evaluate the performance and computational efficiency of the EVA approach for estimation and inference in GLLVMs, compared to other approximate likelihood-based methods. In particular, we compared EVA against standard VA (when this is available) and the Laplace approximation (LA), both of which are already available in the package $\mathsf{gllvm}$ \citep{Nikuetal:2019b}. The simulation setups to be presented were adapted from those previously proposed by \citet{Nikuetal:2019}.


Two main simulation settings were considered, where in the first setting we generated multivariate data with four possible response types (overdispersed counts, binary responses, semi-continuous, and proportional responses), all based on GLLVMs fitted to the testate amoebae data \citep{Dazaseccoetal:2018} considered in Section \ref{sec:example}. Briefly, the true GLLVMs 
included two environmental predictors from the testate amoebae data, and we simulated multivariate data such that 
the number of observational units $n$ increased incrementally while the number of species $m$ stays fixed (consistent with $m/n$ being relatively small in this dataset). More details of this setting as well as the simulation results are presented later on.

In the second setting, we generated multivariate data again with the above four possible response types, but this time based on GLLVMs fitted to a dataset containing species of bird species recorded across Borneo \citep{cleary2005associations}. Unlike in the first simulation setting, here we increased the number of species $m$ while keeping the number of observational units $n$ fixed; this was consistent with the original data itself, where $m/n$ was relatively large). For brevity, we provide details of the simulation design as well the simulation results Appendix \ref{appendix:birdsim}, but do provide a brief discussion of the results later on in the main text. Overall, we see that the two simulations aim to assess the performance of EVA in a variety of settings where the ratio $n/m$ varied from small to larger, and the response type varies.

To assess performance, all methods were in terms of: 1) the empirical bias and the empirical root mean squared error (RMSE) of the regression coefficients and the dispersion parameters (if appropriate), where the averaging is across both the number of simulated datasets as well as across the $m$ species; 2) the corresponding empirical coverage probability of $95\%$ Wald confidence interval (CI), again averaged across both the number of simulated datasets as well as across the $m$ species; 3) the {Procrustes error} between the predicted and true $n \times p$ matrices of latent variables, and similarly between the estimated and true $m \times p$ loading matrices. The Procrustes error is a commonly used measure of evaluating the accuracy of ordinations \citep{peres2001well}; 4) average computation time in seconds. 



\subsection{Setting 1} \label{subsec:setting1}


Multivariate data were simulated from GLLVMs fitted to the testate amoebae data detailed in Section \ref{sec:example} as follows. The original testate amoebae data consisted originally of (overdispersed) count records of $m = 48$ species at $n = 263$ sites across Finland. All GLLVMs fitted to the data included $q = 2$ environmental covariates (water pH and temperature) and $p = 2$ latent variables. No row effects $\alpha_i$ were included. Using the parameter estimates from these GLLVMs as true parameter values (and hence a true data generation mechanism), we then simulated datasets with differing numbers of observational units $n$ while keeping the number of responses fixed to the original size, i.e., $m = 48$. This was accomplished by randomly subsampling rows from the original data and the predicted matrix latent variables. We varied the number of units as $n=50, 120, 190$ and $260$, noting that the full dataset contained 263 sites. We simulated 1000 datasets for each value of $n$.


Datasets with four possible response types were generated, following Sections \ref{subsec:nbgllvm} to \ref{subsec:beta}.
 
\begin{enumerate}
\item Overdispersed counts simulated from a negative binomial GLLVM with log link function fitted to the original testae amoebae count data using the standard VA approach. For each simulated dataset, we then compared negative binomial GLLVMs fitted using EVA to those fitted using standard VA and LA. 

\item Binary responses simulated from a binary GLLVM with either the probit or logit link fitted to a presence-absence version of the original testae amoebae data (formed by setting all positive counts to one, while keeping zero counts at zero). The binary GLLVM was fitted using the standard VA approach (probit link) or the LA approach (logit link). For each simulated dataset, we then compared binary GLLVMs fitted using EVA to those fitted standard VA and/or LA. Standard VA was excluded from the simulations involving a logit link binary GLLVM; see the discussion in Section \ref{subsec:binary}.

\item Semi-continuous responses simulated from a Tweedie GLLVM with log link fitted to the original testate amoebae count data set (after square root transform of the counts) using the LA approach. For each simulated dataset, we then compared Tweedie GLLVMs fitted using EVA to those fitted using LA. Standard VA was again excluded this setting; see the discussion in Section \ref{subsec:tweedie}.

\item Proportions data simulated from a beta GLLVM using the logistic link. As true parameter values for this true model, we used the parameters of the binary GLLVM with logistic link fitted to the presence-absence version of the original testae amoebae data discussed above. Additionally, the true values of the response-specific dispersion parameters $\phi_j$ were drawn independently from the uniform distribution $\text{Unif}(1,3)$. For each simulated dataset, we then compared beta GLLVMs fitted using EVA to those fitted using LA. Standard VA was again excluded this setting; see the discussion in Section \ref{subsec:beta}.
\end{enumerate}

\subsection{Simulation results} 
\label{subsec:simresults}

We discuss simulation results for each of the four response types separately. First, for multivariate overdispersed count data, we first observe the substantial differences in mean computation times between LA and the two variational approximation methods (Table \ref{tab:ame_nb_new}). In fact, EVA was the fastest approach across all four values of $n$ considered. In terms of empirical bias and MSE for the regression coefficients, both EVA and LA tended to perform more similarly to each other, although both were performing slightly poorer than standard VA. Similar trends were also observed with the Procrustes error of the loadings and predicted latent values, while there was little difference between the three methods in terms of coverage probability of the regression coefficients. The differences between all three methods tended to diminish with increasing $n$, although in further exploration we did notice that a non-negligible number of fits produced by LA failed to converge in the sense of not producing a finite log-likelihood value for larger values of $n$ (but this did not appear to hinder its performance much).
Finally, note that the full results including those for the response-specific intercepts $\beta_{0j}$ and dispersion parameters $\phi_j$ can be found in Appendix \ref{appendix:ame_b0_phi}, while in the main text we focus on performance for the regression coefficients relating to the two environmental predictors as these are generally of more interest in practice.

\begin{table*}[htb]
\centering
\small
\begin{tabular}{lllrrrrrrrr}
  \toprule[1.5pt]
 &  &  & \multicolumn{2}{c}{Bias} & \multicolumn{2}{c}{RMSE} & \multicolumn{2}{c}{Coverage} & \multicolumn{2}{c}{Procrustes Err.} \\
 \cmidrule(lr){4-5} \cmidrule(lr){6-7} \cmidrule{8-9} \cmidrule(lr){10-11}
 $n$ & & Time & pH & Temp. & pH & Temp. & pH & Temp. & LV & Loadings \\ 
  \midrule
 $ 50 $ & EVA & 7.69 & 0.015 & -0.136 & 0.644 & 0.677 & 0.938 & 0.935 & 0.270 & 0.531 \\  
    & VA & 7.69 & -0.064 & -0.076 & 0.478 & 0.505 & 0.910 & 0.907 & 0.227 & 0.320 \\ 
   & LA & 70.21 & -0.002 & -0.124 & 0.602 & 0.637 & 0.941 & 0.938 & 0.266 & 0.485 \\  \midrule
  $ 120 $ & EVA & 22.79 & 0.010 & -0.037 & 0.332 & 0.321 & 0.963 & 0.961 & 0.196 & 0.334 \\ 
   & VA & 24.37 & -0.022 & -0.023 & 0.280 & 0.272 & 0.963 & 0.961 & 0.186 & 0.185 \\
   & LA & 143.53 & 0.007 & -0.036 & 0.327 & 0.316 & 0.963 & 0.962 & 0.198 & 0.300 \\  \midrule
  $ 190 $ & EVA & 47.16 & -0.011 & -0.011 & 0.235 & 0.238 & 0.957 & 0.966 & 0.187 & 0.250 \\ 
   & VA & 51.78 & -0.027 & -0.007 & 0.220 & 0.210 & 0.956 & 0.970 & 0.184 & 0.140 \\
   & LA & 252.62 & -0.010 & -0.009 & 0.234 & 0.236 & 0.957 & 0.967 & 0.188 & 0.210 \\  \midrule
  $ 260 $ & EVA & 80.77 & -0.015 & -0.003 & 0.206 & 0.201 & 0.962 & 0.966 & 0.179 & 0.187 \\
   & VA & 89.79 & -0.031 & -0.001 & 0.195 & 0.183 & 0.961 & 0.968 & 0.176 & 0.109 \\
   & LA & 400.90 & -0.015 & -0.001 & 0.208 & 0.204 & 0.960 & 0.962 & 0.180 & 0.175 \\
   \bottomrule[1.5pt]
\end{tabular}
\caption{\label{tab:ame_nb_new} Results from simulation setting 1 involving negative binomial GLLVMs. Performance was assessed in terms of computation time, empirical biases and RMSEs of estimated coefficients for water pH and temperature, empirical coverage probability if 95\% Wald intervals for water pH and temperature, and Procrustes errors of the latent variables and loadings. A trimming factor of $2 \%$ was used when calculating the average biases and RMSEs. 
}
\end{table*}

Next, results for Bernoulli responses involving binary GLLVMs with the probit and logit link are presented in Tables \ref{tab:ame_bin} and \ref{tab:ame_logit}, respectively. Similar to the case with overdispersed counts, EVA was the fastest of the three methods compared, followed by standard VA (when possible) LA, the latter of which was substantially slower. For the probit link case, across all measures of performance and values of $n$ standard VA performed the best, followed  by EVA and then LA, with the latter being rather unstable at the smallest number of observational units considered. It should be noted though that EVA had slight undercoverage in the 95\% CIs, and this persisted even at larger values of $n$. the larger sample sizes. 
Interestingly, for the logit link case where standard VA is not available, EVA outperformed LA across most metrics of comparison and consistently across all values of $n$ considered. EVA was also substantially faster computationally than LA. 

\begin{table*}[htb]
\centering
\small
\begin{tabular}{llrrrrrrrrr}
  \toprule[1.5pt]
  & & & \multicolumn{2}{c}{Bias} & \multicolumn{2}{c}{RMSE} & \multicolumn{2}{c}{Coverage} & \multicolumn{2}{c}{Procrustes} \\
    \cmidrule(lr){4-5} \cmidrule(lr){6-7} \cmidrule(lr){8-9} \cmidrule(lr){10-11}
  $n$ & & Time & pH & Temp. & pH & Temp. & pH & Temp. & LV & Loadings \\ 
  \midrule
$50$ & EVA & 1.84 & -0.036 & -0.063 & 0.247 & 0.256 & 0.965 & 0.963 & 0.363 & 0.372 \\
   & VA & 3.06 & -0.044 & -0.051 & 0.279 & 0.276 & 0.979 & 0.973 & 0.269 & 0.286 \\
   & LA & 99.76 & 0.576 & 0.466 & 6.264 & 4.657 & 0.979 & 0.973 & 0.320 & 0.869 \\   \midrule
  $120$ & EVA & 7.33 & -0.008 & -0.005 & 0.201 & 0.186 & 0.923 & 0.930 & 0.276 & 0.211 \\  
   & VA & 9.60 & -0.002 & -0.011 & 0.151 & 0.140 & 0.968 & 0.973 & 0.242 & 0.147 \\ 
   & LA & 127.98 & 0.090 & 0.026 & 0.397 & 0.301 & 0.968 & 0.973 & 0.264 & 0.796 \\  \midrule
  $190$ & EVA & 16.36 & 0.013 & 0.007 & 0.150 & 0.149 & 0.913 & 0.930 & 0.291 & 0.169 \\  
   & VA & 21.70 & 0.000 & 0.002 & 0.115 & 0.111 & 0.956 & 0.967 & 0.250 & 0.111 \\ 
   & LA & 139.94 & 0.011 & 0.016 & 0.136 & 0.140 & 0.956 & 0.967 & 0.269 & 0.547 \\  \midrule
  $260$ & EVA & 29.93 & -0.009 & -0.002 & 0.136 & 0.129 & 0.920 & 0.919 & 0.266 & 0.129 \\
   & VA & 39.27 & -0.013 & -0.006 & 0.100 & 0.094 & 0.960 & 0.966 & 0.233 & 0.082 \\ 
   & LA & 168.07 & -0.004 & 0.005 & 0.114 & 0.110 & 0.960 & 0.966 & 0.245 & 0.313 \\
   \bottomrule[1.5pt]
\end{tabular}
\caption{\label{tab:ame_bin} Results from simulation setting 1 involving binary GLLVMs with the probit link. Performance was assessed in terms of computation time, empirical biases and RMSEs of estimated coefficients for water pH and temperature, empirical coverage probability if 95\% Wald intervals for water pH and temperature, and Procrustes errors of the latent variables and loadings. A trimming factor of $2 \%$ was used when calculating the average biases and RMSEs.
}
\end{table*}

\begin{table*}[tb]
\centering
\small
\begin{tabular}{llrrrrrrrrr}
  \toprule[1.5pt]
  & & & \multicolumn{2}{c}{Bias} & \multicolumn{2}{c}{RMSE} & \multicolumn{2}{c}{Coverage} & \multicolumn{2}{c}{Procrustes} \\
    \cmidrule(lr){4-5} \cmidrule(lr){6-7} \cmidrule(lr){8-9} \cmidrule(lr){10-11}
  $n$ & & Time & pH & Temp. & pH & Temp. & pH & Temp. & LV & Loadings \\ 
  \midrule
$50$ & EVA & 1.17 & -0.187 & -0.034 & 1.022 & 0.927 & 0.957 & 0.958 & 0.275 & 0.538 \\
   & LA & 47.44 & 1.541 & 2.225 & 17.464 & 13.022 & 0.851 & 0.883 & 0.286 & 0.705 \\    \midrule
  $120$ & EVA & 4.62 & -0.006 & 0.048 & 0.464 & 0.406 & 0.945 & 0.963 & 0.237 & 0.496 \\  
   & LA & 118.89 & 2.798 & 1.416 & 4.397 & 2.883 & 0.936 & 0.955 & 0.233 & 0.774 \\  \midrule
  $190$ & EVA & 9.70 & 0.006 & 0.066 & 0.328 & 0.307 & 0.933 & 0.958 & 0.235 & 0.440 \\  
   & LA & 172.64 & 0.993 & 0.358 & 1.513 & 0.986 & 0.942 & 0.959 & 0.235 & 0.727 \\   \midrule
  $260$ & EVA & 17.36 & -0.030 & 0.037 & 0.282 & 0.241 & 0.929 & 0.957 & 0.217 & 0.413 \\
   & LA & 227.07 & 0.369 & 0.112 & 0.712 & 0.437 & 0.947 & 0.959 & 0.217 & 0.662 \\
   \bottomrule[1.5pt]
\end{tabular}
\caption{\label{tab:ame_logit} Results from simulation setting 1 involving binary GLLVMs with the logit link. Performance was assessed in terms of computation time, empirical biases and RMSEs of estimated coefficients for water pH and temperature, empirical coverage probability if 95\% Wald intervals for water pH and temperature, and Procrustes errors of the latent variables and loadings. A trimming factor of $2 \%$ was used when calculating the average biases and RMSEs.
}
\end{table*}

Finally, turning to semi-continuous data with Tweedie GLLVMs and proportions data with beta GLLVMs, noting that standard VA is again not available for these two response distributions, the performance of EVA and LA in both cases were fairly similar across all metrics and across the four values of $n$ considered (see Table \ref{tab:ame_tweedie_new} for the Tweedie GLLVM case and Appendix \ref{appendix:ame_b0_phi} for the beta GLLVM case). EVA was substantially faster and scaled computationally better than LA. 

\begin{table*}[tb]
\centering
\small
\begin{tabular}{llrrrrrrrrr}
  \toprule[1.5pt]
 & &   & \multicolumn{2}{c}{Bias} & \multicolumn{2}{c}{RMSE} & \multicolumn{2}{c}{Coverage} & \multicolumn{2}{c}{Procrustes} \\
 \cmidrule(lr){4-5} \cmidrule(lr){6-7} \cmidrule(lr){8-9} \cmidrule(lr){10-11}
 $n$ & & Time & pH & Temp.  & pH & Temp. & pH & Temp. & LV & Loadings \\ 
  \midrule
 $50$ & EVA & 12.64  & -0.072 & -0.086 & 0.483 & 0.474 & 0.971 & 0.967 & 0.221 & 0.382 \\ 
   & LA & 81.72  & -0.078 & -0.085 & 0.466 & 0.462 & 0.971 & 0.967 & 0.222 & 0.362 \\    \midrule
  $120$ & EVA & 34.60  & -0.001 & -0.014 & 0.254 & 0.234 & 0.970 & 0.973 & 0.199 & 0.269 \\  
   & LA & 180.33  & -0.005 & -0.015 & 0.248 & 0.230 & 0.969 & 0.972 & 0.197 & 0.237 \\   \midrule
  $190$ & EVA & 67.48  & -0.005 & 0.009 & 0.189 & 0.177 & 0.957 & 0.972 & 0.199 & 0.210 \\
   & LA & 296.03  & -0.008 & 0.009 & 0.187 & 0.175 & 0.956 & 0.972 & 0.199 & 0.181 \\  \midrule
  $260$ & EVA & 109.15  & -0.013 & 0.009 & 0.166 & 0.151 & 0.961 & 0.968 & 0.189 & 0.172 \\  
   & LA & 424.89  & -0.016 & 0.009 & 0.164 & 0.150 & 0.960 & 0.968 & 0.189 & 0.144 \\
   \bottomrule[1.5pt]
\end{tabular}
\caption{\label{tab:ame_tweedie_new} Results from simulation setting 1 involving Tweedie GLLVMs. Performance was assessed in terms of computation time, empirical biases and RMSEs of estimated coefficients for water pH and temperature, empirical coverage probability if 95\% Wald intervals for water pH and temperature, and Procrustes errors of the latent variables and loadings. A trimming factor of $2 \%$ was used when calculating the average biases and RMSEs.
}
\end{table*}

Overall, the results from simulation setting 1 demonstrate that in terms of estimation and inferential accuracy, the performance of EVA for GLLVMs is competitive with that of LA, and similar to that of VA (which tended to be the best performer overall) for larger number of observational units $n$. EVA is, on average, the most computationally efficient method of the three considered. Given its more universal applicability, this suggests that EVA may be a more suitable choice in scenarios particularly when standard VA cannot be applied. 
Results from simulation setting 2, which are provided in Appendix \ref{appendix:birdsim}, largely support these conclusions, with the one major difference between that with fixed $n$ and increasing number of responses $m$, the differences in computational times between the three methods are even more dramatic, and EVA becomes even more scalable compared to LA and standard VA.

\section{Application to testate amoebae data}
\label{sec:example}

We illustrate the application of EVA for fitting a negative binomial GLLVM (using the log link function) to a multivariate abundance dataset of testate amoebae available from \citet{Dazaseccoetal:2018}. Note the data was used as the basis for simulation setting 1 in Section \ref{subsec:setting1}. The testate amoebae dataset consisted of counts from $m=48$ testate amoebae species measured on $n = 248$ peatland sites spread throughout Finland. In addition to water pH and temperature, the dataset also contained a factor variable on the type of land use for each site (forestry, natural and restored). 

\begin{figure}[htb!]
    \centering
    \includegraphics[height = 0.4\textheight]{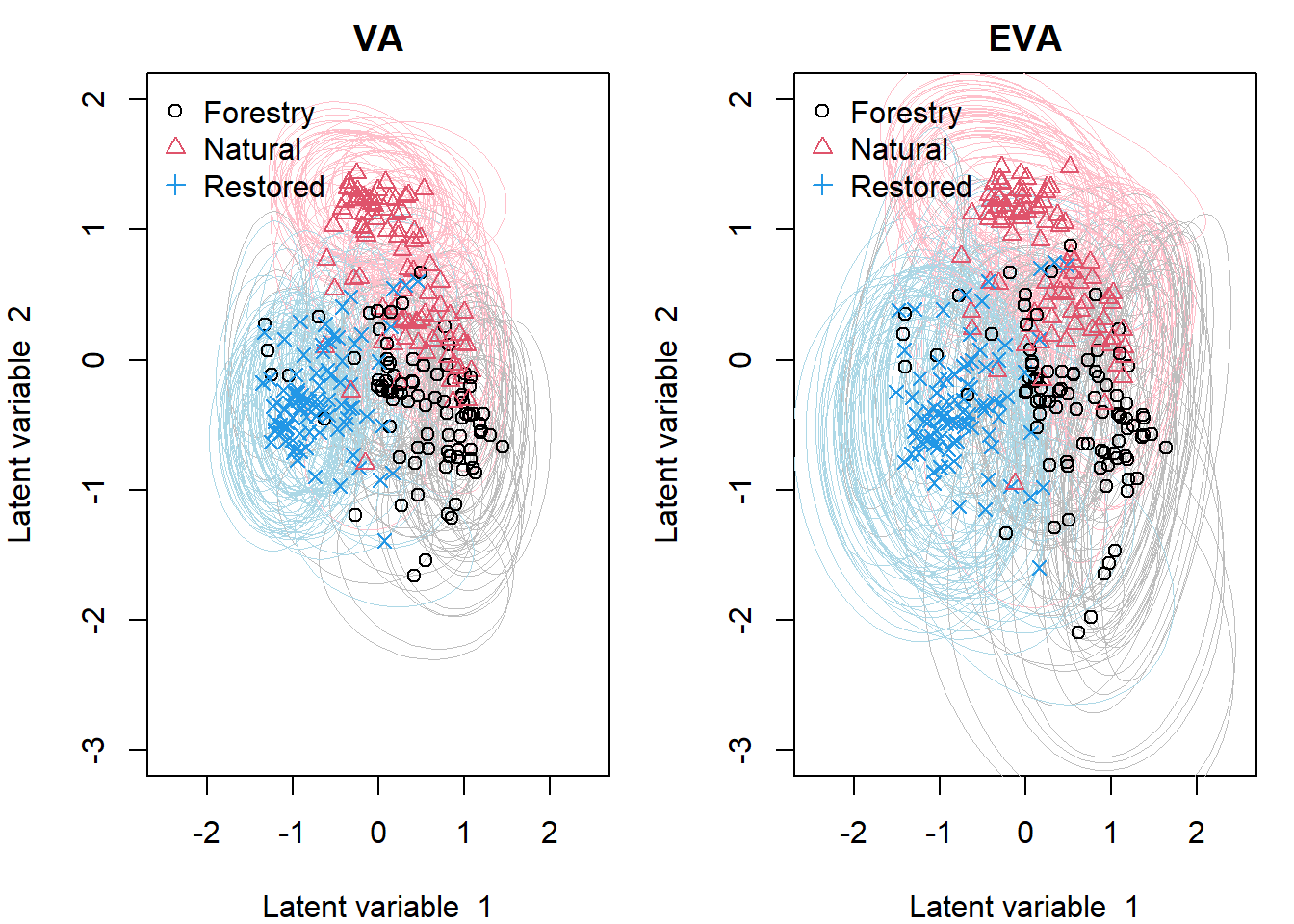}
    \includegraphics[height = 0.4\textheight]{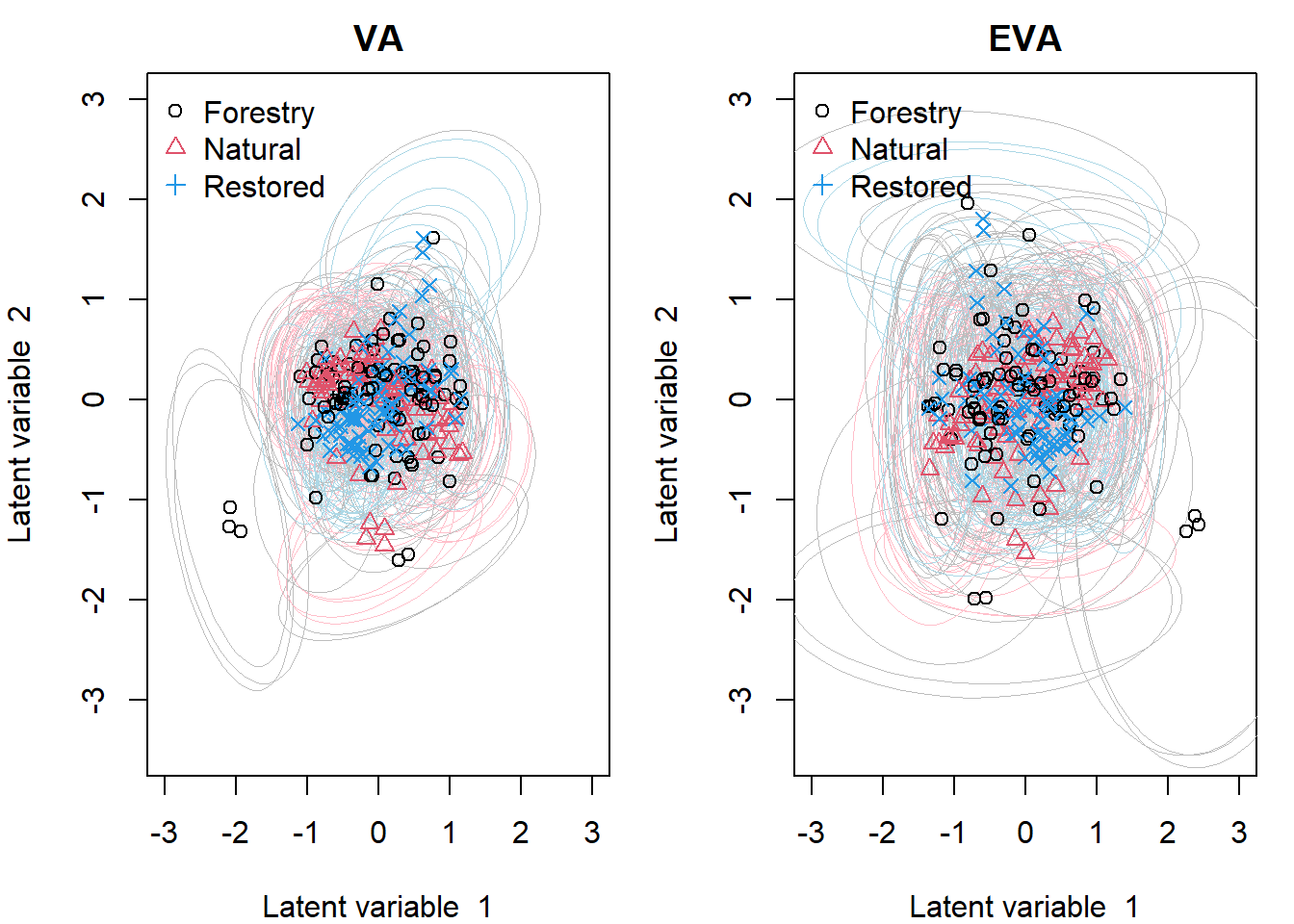}
    \caption{Model-based unconstrained ordination (top row) and residual ordination (bottom row) of the sites in the testate amoebae data, along with 95\% CMSEP-based prediction regions. The ordinations are constructed based on fitting an negative binomial GLLVMs using standard VA (left column) and EVA (right column), where in the top row no covariates are included while in the bottom row the covariates water pH, temperature, and land use were included. The sites are colored and marked according to their type of land use.}
    \label{fig:ordiplot_amoeba}
\end{figure}

We began by fitting a negative binomial GLLVM using EVA assuming $p = 2$ latent variables, with no covariates or row effects included, as a means of model-based unconstrained ordination. The aim here was to assess whether the sites tended to cluster according to land usage, as based on their predicted latent variable scores. For comparison, we also fitted the same GLLVM using standard VA to assess if there were any noticeable differences in the conclusions and inferences drawn between the two methods of estimation. Indeed, from the previous simulation results we observed that if the number of observational units $n$ was sufficiently large then there was comparably little difference between EVA and standard VA, although the former was tended to slightly faster for fitting. The top row of Figure \ref{fig:ordiplot_amoeba} presents the resulting unconstrained ordination plots using both standard VA (left panel) and EVA (right panel). The ordinations of the sites given by the two methods match each other quite well, and there is evidence that the sites are clustered according to their on the land usage type. There is generally much less uncertainty in the prediction regions resulting from standard VA compared to EVA. 

\FloatBarrier 

Based on the above results, we proceeded to fit a negative binomial GLLVM but this time including water pH, temperature, and land use type (as a factor with dummy variables) as covariates. The resulting \emph{residual} ordination plots, which may be interpreted as a visualization of residual covariation between species after accounting for the measured covariates \citep{warton2015so,Bjorketal} are presented in the bottom row of Figure \ref{fig:ordiplot_amoeba}. Not surprisingly, 
after controlling for the land use, the sites exhibit a much more random pattern (using both EVA and standard VA), and on the whole are more closely clustered together compared to the unconstrained ordination plot. The prediction regions produced by EVA are again noticeably bigger than those produced by standard VA.

By looking at the definition of CMSEP in Section \ref{sec:inference}, one may hypothesize the larger prediction regions resulting from using EVA seen in the ordinations may be, at least partially, a consequence of the elements of the estimated variational covariance matrices $\bm{\hat A_i}$ being larger. In Appendix \ref{appendix:example}, we provide some results that supports this this idea, namely that the traces of the variational covariance matrices from EVA are typically greater than those produced by standard VA. These larger traces and larger predictions regions as a whole are not overly surprising, since EVA uses a Taylor approximation rather than the exact form of conditional distribution of the responses. 

On the other hand, when we examine the amount of covariation within and between species that is explained by the covariates, as quantified by calculating the relative change in the trace of the estimated residual covariance matrix $\hat{\bm \Sigma} = \hat{\bm \Gamma} \hat{\bm \Gamma}^\intercal$, we observe that, according to models fitted using EVA, water pH and temperature together explain only $15.5 \%$ of the covariation in the GLLVM, whereas when land usage is also included this rose to $39.9 \%$. For models fitted using standard VA, these percentages were $15.8 \%$ and $38.7 \%$, respectively. The fact that these percentages were fairly similar between models fitted EVA versus standard VA provides some reassurance of the inferences and conclusions obtained the former. This result is further supported when we examine plots of the estimated regression coefficients and corresponding 95\% Wald intervals from both fits. These plots for pH, temperature and land usage are presented in Appendix \ref{appendix:example}, from which we see that the conclusions produced by EVA and VA are almost exact match, with the list of covariate effects deemed statistically significant differing only by few.

\section{Discussion}\label{sec:discussion}

In this article, we have proposed the method of extended variational approximations (EVA) for fast and universal fitting of GLLVMs. EVA builds on the ongoing research into variational approximations for GLLVMs, but broadens to allow for any combination of (parametric) response distribution and link function to be used. Based on extensive simulation studies, the performance accuracy of EVA lies somewhere between the standard method of VA, and the well-known method of Laplace approximation (LA). This, combined with strong computational efficiency and scalability (in fact, even more so even standard VA) suggests that EVA presents an exciting and potentially instrumental avenue of further research into computationally efficient estimation and inference for GLLVMs as a whole. Indeed, the EVA approach is potentially more straightforward to derive and implement for more advanced types of GLLVMs such as those involving temporally or spatially dependent latent variables \citep{Ovaskainenetal:2017}, detection probabilities \citep{tobler2019joint}, and GLLVMs coupled with sparse regularization penalties for variable selection \citep{hui2018order}. One particularly important avenue of future use for EVA is as a computationally efficient way to fit mixed-response GLLVMs, when the columns of $\bm Y$ correspond to different types of response variables \citep{Sammeletal:1997}, and research is currently being done to implement this as part of the \texttt{gllvm} package. 

From a theoretical standpoint, the derivation of the EVA log-likelihood leaves a fair amount on room for modification, related most notably to the use of the Taylor approximation for the conditional distribution of the response, $\log f(\bm y | \bm u, \bm \Psi)$. For instance, \citet{wang2013variational} explored a variant 
where the Taylor approximation is taken to be the point which maximizes $\log f(\bm y | \bm u, \bm \Psi)$ with respect to the latent variables $\bm{u}$. This choice leads to yet another method which authors called \textit{Laplace variational inference}. Yet another choice is to center the Taylor approximation around the mean of the variational distribution from the previous \emph{iterative} step of the optimization algorithm, although according to the authors this approach often did not lead to desirable results in terms of model convergence. To our knowledge, Laplace variational inference has not been considered for GLLVMs, and its performance in comparison to EVA is an avenue of future research, although we conjecture that Laplace variational inference might be computationally slower due to the additional burden of finding an additional maxima point. Finally, the effect of using higher order Taylor expansions could be explored, along with developing general large sample properties for EVA for GLLVMs. 

\section*{Acknowledgements}
FKCH was supported by an Australian Research Council Discovery Early Career Research Award. PK and ST were supported by the Kone foundation and JN was supported by the Maj and Tor Nessling foundation.


\appendix
\section{Proofs and derivations} 
\label{appendix:proofs}

\setcounter{figure}{0} \renewcommand{\thefigure}{\Alph{section}\arabic{figure}}
\setcounter{table}{0} \renewcommand{\thetable}{\Alph{section}\arabic{table}}

\subsection{Proof of Theorem 1}
The first two terms in EVA log-likelihood are clear by looking from $(5)$ in the main text. Thus it suffices to explicitly derive the term $\Tr(\bm{H}_i(\bm{a}_i, \bm \Psi) \bm{A}_i)$.  We have
\begin{align*}
    \frac{\partial \log f(y_{ij}|\bm{u}_i, \bm \Psi)}{\partial \bm{u}_i} &= \frac{\partial}{\partial \bm{u}_i}\left\{ h_j(y_{ij})b_j\{g^{-1}(\eta_{ij})\} - c_j\{g^{-1}(\eta_{ij})\} + d_j(y_{ij}) \right\} \\
    &= h_j(y_{ij})b_j'\{g^{-1}(\eta_{ij})\}(g^{-1})'(\eta_{ij}) \bm \lambda_j^\intercal - c_j'\{g^{-1}(\eta_{ij})\}(g^{-1})'(\eta_{ij})\bm \lambda_j^\intercal.
\end{align*}    
The second derivative then follows as 
\begin{align*}
\frac{\partial^2 \log f(y_{ij}|\bm{u}_i, \bm \Psi)}{\partial \bm{u}_i\partial \bm{u}_i^\intercal} &= h_j(y_{ij}) \left\{b_j''\{g^{-1}(\eta_{ij})\} \{(g^{-1})'(\eta_{ij}) \}^2 + b_j'\big\{g^{-1}(\eta_{ij})\}(g^{-1})''(\eta_{ij}) \right\} \bm \lambda_j \bm \lambda_j^\intercal \\
    &\quad - \left\{c_j''\{g^{-1}(\eta_{ij})\} \{(g^{-1})'(\eta_{ij})\}^2 + c_j'\{ g^{-1}(\eta_{ij})\}g^{-1})''(\eta_{ij}) \right\} \bm \lambda_j \bm \lambda_j^\intercal.
    \intertext{By using the inverse function rule of differentiation, this can be simplified into}
    \frac{\partial^2 \log f(y_{ij}|\bm{u}_i, \bm \Psi)}{\partial \bm{u}_i\partial \bm{u}_i^\intercal} &= \left\{\frac{h_j(y_{ij})b_j''(\mu_{ij}) - c_j''(\mu_{ij})}{\{g'(\mu_{ij}) \}^2} - \frac{h_j(y_{ij})b_j'(\mu_{ij})-c_j'(\mu_{ij})}{\{g'(\mu_{ij}) \}^3/g''(\mu_{ij})} \right\}\bm \lambda_j \bm \lambda_j^\intercal.
\end{align*}
Applying the above result together with the fact that $$\Tr(\bm\lambda_j \bm \lambda_j^\intercal \bm{A}_i) = \Tr(\bm \lambda_j^\intercal \bm{A}_i \bm \lambda_j) = \bm \lambda_j^\intercal \bm{A}_i \bm \lambda_j,$$ completes the proof.

\subsection{Proof of Corollary 1}
For some fixed $i \in \{1,\dots,n\}$ and $j \in \{1,\dots,m\}$, denote $h = h_j, b = b_j, c=c_j, \mu = \mu_{ij}$ and $y = y_{ij}$. Then the multiplier in the summand in the third term of expression from Theorem $1$ becomes
\begin{align*}
    &\frac{h(y)b''(\mu)-c''(\mu)}{\{g'(\mu)\}^2} - \frac{h(y)b'(\mu)-c'(\mu)}{\{g'(\mu) \}^3/g''(\mu)} = \frac{h(y)b''(\mu)-c''(\mu)}{\{b'(\mu)\}^2} - \frac{h(y)b'(\mu)-c'(\mu)}{\{b'(\mu) \}^3/b''(\mu)} \\
    &= \frac{\frac{b'(\mu)}{b''(\mu)} \left\{h(y) b''(\mu) - c''(\mu)\right\} - h(y)b'(\mu) + c'(\mu)}{\{b'(\mu) \}^3/b''(\mu)} \\
    &= \frac{b'(\mu)h(y)b''(\mu)-b'(\mu)c''(\mu) - b'(\mu)h(y)b''(\mu) + b''(\mu) c'(\mu)}{\{b'(\mu)\}^3} 
    = \frac{b''(\mu)c'(\mu) - b'(\mu)c''(\mu)}{\{b'(\mu)\}^3},
\end{align*}
which proves the result.

\subsection{Derivations for beta and Tweedie GLLVMs}
\label{appendix:derivations}

\subsubsection*{Tweedie model}
Assume that conditional on the latent variables, the responses $y_{ij}$ follow a Tweedie distribution with response-specific dispersion parameters $\phi_j > 0$ and response-common power parameter $1 < \nu < 2$. The log-density of the Tweedie distribution in given Section $4.3$ of the main text, with the generalized Bessel function in the normalizing constant taking the form 
\begin{align*}
W(y_{ij}, \phi_j, \nu) = \sum_{k=1}^\infty \frac{y_{ij}^{-k\gamma} (\nu-1)^{k\gamma}}{\phi_j^{k(1-\gamma)}(2-\nu)^k k!\Gamma(-k\gamma)},
\end{align*}
where $\gamma = (2-\nu)/(1-\nu)$. If we assume a Tweedie GLLVM with the log link function, then the second derivatives of $\log f(y_{ij}|{\bo u_i}, \bo \Psi)$ with respect to $\bm{u}_i$, are 
\begin{align*}
\frac{\partial^2 \log f(y_{ij}|{\bo u_i}, \bo \Psi)}{\partial \bo u_i \partial \bo u_i^\intercal} &= \left\{-\frac{1}{\phi_j}(2-\nu)\exp(2\eta_{ij} - \nu\eta_{ij})\right\} \bo\lambda_j \bo\lambda_j^\intercal, \\ \intertext{when $y_{ij}=0$, and}
\frac{\partial^2 \log f(y_{ij}|{\bo u_i}, \bo \Psi)}{\partial \bo u_i \partial \bo u_i^\intercal} &= \frac{1}{\phi_j}\left\{y_{ij}(1-\nu)\exp(\eta_{ij}-\nu\eta_{ij}) - (2-\nu)\exp(2\eta_{ij}-\nu\eta_{ij} )   \right\} \bo \lambda_j \bo \lambda_j^\intercal,
\end{align*}
when $y_{ij} > 0$. By plugging these derivatives into equation (5) of the main text and following some straightforward algebra, the EVA log-likelihood function for the Tweedie GLLVM can be obtained.

\subsubsection*{Beta model}
Consider a beta GLLVM where the log-density of the beta distribution is given in Section $4.4$ of the main text. Suppose we use the logit link function, such that $\mu_{ij} = \exp(\eta_{ij})/(\exp(\eta_{ij})+1)$. Then by applying straightforward algebra we can obtain the second derivatives of $\log f(y_{ij}|{\bo u_i}, \bo \Psi)$ with respect to $\bm{u}_i$ as 
\begin{align*}
    \frac{\partial^2 \log f(y_{ij}|{\bm{u}_i}, \bm \Psi)}{\partial \bm{u}_i \partial \bm{u}_i^\intercal} &= \left\{-\psi_1(\mu_{ij}\phi_j)\phi_j^2(\mu_{ij}')^2 - \phi_j\mu_{ij}''\psi(\mu_{ij}\phi_j)\right\} \bm \lambda_j \bm \lambda_j^\intercal \nonumber \\
    &\quad + \left\{- \psi_1\big\{(1-\mu_{ij})\phi_j\big\}\phi_j(\mu_{ij}')^2 + \psi\{(1-\mu_{ij})\phi_j\}\phi_j\mu_{ij}'' \right\} \bm \lambda_j \bm \lambda_j^\intercal \nonumber\\[1ex]
    &\quad+ \big[\phi_j\mu_{ij}''\big\{\log(y_{ij}) - \log(1-y_{ij})\big\}\big] \bm \lambda_j \bm \lambda_j^\intercal, 
\end{align*}
where $\psi(\cdot)$ and $\psi_1(\cdot)$ are the digamma and trigamma functions respectively, $\mu_{ij}'$ is a shorthand notation for $\mu'(\eta_{ij}) = \exp(\eta_{ij})/(\exp(\eta_{ij})+1)^2$, and similarly $\mu_{ij}''$ is a shorthand notation for $\mu''(\eta_{ij}) = \{\exp(3\eta_{ij}) - \exp(\eta_{ij})\}/(\exp(\eta_{ij})+1)^4$. By plugging these derivatives into equation $(5)$ of the main text, the EVA log-likelihood function for the beta GLLVM can be obtained.

\section{Additional results for simulation studies} \label{appendix:simu}

\subsection{Setting 1}
\label{appendix:ame_b0_phi}

Table \ref{tab:ame_beta_new} presents the results from simulation setting 1 for the beta GLLVM. As discussed in the main text, the performance of EVA and LA was similar across all metrics and across the four values of $n$ considered. However, EVA was substantially faster and scaled computationally better than LA. 

\begin{table*}[htb]
\centering
\begin{tabular}{llrrrrrrrrr}
  \toprule[1.5pt]
 & &   & \multicolumn{2}{c}{Bias} & \multicolumn{2}{c}{RMSE} & \multicolumn{2}{c}{Coverage} & \multicolumn{2}{c}{Procrustes} \\
 \cmidrule(lr){4-5} \cmidrule(lr){6-7} \cmidrule(lr){8-9} \cmidrule(lr){10-11}
 $n$ & & Time & pH & Temp.  & pH & Temp. & pH & Temp. & LV & Loadings \\ 
  \midrule
 $50$ & EVA & 30.09 & -0.041 & -0.047 & 0.155 & 0.175 & 0.932 & 0.925 & 0.103 & 0.180 \\
   & LA & 161.59 & -0.039 & -0.046 & 0.150 & 0.173 & 0.933 & 0.924 & 0.097 & 0.176 \\    \midrule
  $120$ & EVA & 110.23 & -0.007 & -0.015 & 0.103 & 0.096 & 0.927 & 0.929 & 0.082 & 0.130 \\    
   & LA & 598.81 & -0.006 & -0.015 & 0.102 & 0.095 & 0.928 & 0.930 & 0.081 & 0.130 \\   \midrule
  $190$ & EVA & 217.37 & -0.009 & 0.001 & 0.100 & 0.080 & 0.901 & 0.929 & 0.088 & 0.124 \\ 
   & LA & 1342.71 & -0.006 & 0.003 & 0.093 & 0.076 & 0.901 & 0.929 & 0.086 & 0.118 \\  \midrule
  $260$ & EVA & 393.64 & -0.026 & -0.013 & 0.094 & 0.077 & 0.921 & 0.920 & 0.088 & 0.126 \\ 
   & LA & 1483.47 & -0.018 & -0.011 & 0.074 & 0.069 & 0.922 & 0.921 & 0.078 & 0.113 \\ 
   \bottomrule[1.5pt]
\end{tabular}
\caption{\label{tab:ame_beta_new} Results from simulation setting 1 involving beta GLLVMs. Performance was assessed in terms of computation time, empirical biases and RMSEs of estimated coefficients for water pH and temperature, empirical coverage probability if 95\% Wald intervals for water pH and temperature, and Procrustes errors of the latent variables and loadings. A trimming factor of $2 \%$ was used when calculating the average biases and RMSEs.
}
\end{table*}

As discussed in the main text, in many applications of GLLVMs estimation and inference on the intercepts $\beta_{0j}$ and dispersion parameters $\phi_j$ tend to be of secondary interest, when compared to the inference on the regression coefficients corresponding to measured covariates. For completeness though, below we present results for these parameters in the case of the four response types considered in simulation setting 1. 

Overall, we found that perhaps not surprisingly, estimation and inference of intercepts and especially the dispersion parameters tended to poses a more relatively more difficult task across the four response types and values of $n$ considered (Tables \ref{tab:ame_nb_aux}-\ref{tab:ame_beta_aux}). For example, based on the results regarding the empirical coverage probabilities of the 95\% Wald intervals, we observe consistent albeit slight undercoverage for the $\phi_j$'s in the negative binomial, Tweedie, and beta GLLVMs. Except at small values of $n$ for the Bernoulli response case, EVA and LA performed similarly across most metrics. 

\begin{table*}[htb]
\centering
\begin{tabular}{llrrrrrr}
  \toprule[1.5pt]
    &    & \multicolumn{2}{c}{Bias} & \multicolumn{2}{c}{RMSE} & \multicolumn{2}{c}{Coverage} \\
    \cmidrule(lr){3-4} \cmidrule(lr){5-6} \cmidrule(lr){7-8}
  $n$ &   & $\beta_{0j}$ & $\phi_j$ & $\beta_{0j}$ & $\phi_j$ & $\beta_{0j}$ & $\phi_j$ \\ 
  \midrule
  $50$ & EVA & -1.278 & -3.916 & 1.439 & 4.408 & 0.931 & 0.604 \\ 
   & VA & -0.363 & -2.905 & 0.576 & 3.538 & 0.921 & 0.721 \\  
   & LA & -1.115 & -3.903 & 1.267 & 4.369 & 0.936 & 0.608 \\   \midrule
  $120$ & EVA & -0.433 & -2.211 & 0.577 & 2.929 & 0.947 & 0.788 \\
   & VA & -0.091 & -1.095 & 0.325 & 2.341 & 0.948 & 0.882 \\ 
   & LA & -0.415 & -2.210 & 0.559 & 2.935 & 0.947 & 0.788 \\  \midrule
  $190$ & EVA & -0.232 & -1.524 & 0.357 & 2.304 & 0.957 & 0.842 \\
   & VA & -0.015 & -0.472 & 0.250 & 1.925 & 0.951 & 0.921 \\
   & LA & -0.231 & -1.552 & 0.355 & 2.347 & 0.956 & 0.840 \\ \midrule
  $260$ & EVA & -0.165 & -1.139 & 0.280 & 1.936 & 0.959 & 0.868 \\
   & VA & 0.023 & -0.106 & 0.219 & 1.693 & 0.943 & 0.932 \\ 
   & LA & -0.172 & -1.237 & 0.286 & 2.003 & 0.959 & 0.863 \\ 
   \bottomrule[1.5pt]
\end{tabular}
\caption{\label{tab:ame_nb_aux} Additional results for from simulation setting 1 involving negative binomial GLLVMs, for response-specific intercepts and overdispersion parameters. Performance was assessed in terms of empirical bias and RMSE, and empirical coverage probability if 95\% Wald intervals. The average bias and RMSE were calculated using a trimming factor of 2 \%. 
}
\end{table*}

\begin{table*}[htb]
\centering
\begin{tabular}{llrrrrrr}
  \toprule[1.5pt]
  & & \multicolumn{2}{c}{Bias} & \multicolumn{2}{c}{RMSE} & \multicolumn{2}{c}{Coverage} \\
    \cmidrule(lr){3-4} \cmidrule(lr){5-6} \cmidrule(lr){7-8} 
  $n$ & & Probit & Logit & Probit & Logit & Probit & Logit \\ 
  \midrule
$50$ & EVA & -0.014 & -1.586 & 0.272 & 2.046 & 0.943 & 0.945 \\
   & VA & -0.130 & - & 0.312 & - & 0.980 & - \\
   & LA & -16.487 & -47.494 & 17.608 & 49.613 & 0.836 & 0.761 \\   \midrule
  $120$ & EVA & -0.176 & -0.487 & 0.292 & 0.818 & 0.890 & 0.934 \\  
   & VA & -0.016 & - & 0.158 & - & 0.964 & - \\ 
   & LA & -0.962 & -13.304 & 1.084 & 13.671 & 0.946 & 0.919 \\  \midrule
  $190$ & EVA & -0.061 & -0.328 & 0.187 & 0.571 & 0.909 & 0.939 \\  
   & VA & 0.001 & - & 0.119 & - & 0.963 & - \\ 
   & LA & -0.119 & -4.695 & 0.197 & 4.915 & 0.962 & 0.940 \\  \midrule
  $260$ & EVA & -0.053 & -0.161 & 0.155 &  0.430 & 0.919 & 0.928\\
   & VA & 0.018 & - & 0.103 & - & 0.955 & - \\ 
   & LA & -0.075 & -1.640 & 0.145 & 1.842 & 0.965 & 0.945 \\
   \bottomrule[1.5pt]
\end{tabular}
\caption{\label{tab:ame_binintercepts} Additional results for from simulation setting 1 involving binary GLLVMs with the probit or logit link functions, for response-specific intercepts. Performance was assessed in terms of empirical bias and RMSE, and empirical coverage probability if 95\% Wald intervals. The average bias and RMSE were calculated using a trimming factor of 2 \%. Note standard VA is not available for the logit link case.
}
\end{table*}

\begin{table*}[tb]
\centering
\begin{tabular}{llrrrrrr}
  \toprule[1.5pt]
    &    & \multicolumn{2}{c}{Bias} & \multicolumn{2}{c}{RMSE} & \multicolumn{2}{c}{Coverage} \\
    \cmidrule(lr){3-4} \cmidrule(lr){5-6} \cmidrule(lr){7-8}
  $n$ &   & $\beta_{0j}$ & $\phi_j$ & $\beta_{0j}$ & $\phi_j$ & $\beta_{0j}$ & $\phi_j$ \\ 
  \midrule
  $50$ &  EVA & -0.787 & -0.114 & 1.009 & 0.252 & 0.973 & 0.874 \\ 
   & LA & -0.706 & -0.112 & 0.932 & 0.250 & 0.973 & 0.876 \\    \midrule
  $120$ & EVA & -0.240 & -0.042 & 0.472 & 0.156 & 0.963 & 0.923 \\
   & LA & -0.204 & -0.041 & 0.442 & 0.155 & 0.963 & 0.924 \\  \midrule
  $190$ & EVA & -0.120 & -0.027 & 0.330 & 0.122 & 0.963 & 0.934 \\
   &LA & -0.100 & -0.025 & 0.317 & 0.121 & 0.963 & 0.935 \\    \midrule
  $260$ & EVA & -0.060 & -0.017 & 0.273 & 0.104 & 0.958 & 0.938 \\ 
   & LA & -0.045 & -0.016 & 0.265 & 0.104 & 0.957 & 0.939 \\
   \midrule
\end{tabular}
\caption{\label{tab:ame_tweedie_aux} Additional results for from simulation setting 1 involving Tweedie GLLVMs, for response-specific intercepts and overdispersion parameters. Performance was assessed in terms of empirical bias and RMSE, and empirical coverage probability if 95\% Wald intervals. The average bias and RMSE were calculated using a trimming factor of 2 \%. 
}
\end{table*}

\begin{table*}[tb]
\centering
\begin{tabular}{llrrrrrr}
  \toprule[1.5pt]
    &    & \multicolumn{2}{c}{Bias} & \multicolumn{2}{c}{RMSE} & \multicolumn{2}{c}{Coverage} \\
    \cmidrule(lr){3-4} \cmidrule(lr){5-6} \cmidrule(lr){7-8}
  $n$ &   & $\beta_{0j}$ & $\phi_j$ & $\beta_{0j}$ & $\phi_j$ & $\beta_{0j}$ & $\phi_j$ \\ 
  \midrule
  $50$ & EVA & 0.042 & 0.312 & 0.196 & 0.483 & 0.925 & 0.942 \\ 
   & LA & 0.039 & 0.311 & 0.189 & 0.477 & 0.925 & 0.941 \\   \midrule
  $120$ & EVA & 0.030 & 0.102 & 0.129 & 0.264 & 0.913 & 0.927 \\
   & LA & 0.030 & 0.102 & 0.128 & 0.263 & 0.914 & 0.928 \\   \midrule
  $190$ & EVA & 0.036 & 0.050 & 0.109 & 0.209 & 0.923 & 0.926 \\ 
   & LA & 0.031 & 0.052 & 0.100 & 0.201 & 0.923 & 0.926 \\   \midrule
  $260$ & EVA & 0.049 & 0.030 & 0.117 & 0.184 & 0.919 & 0.923 \\
   & LA & 0.038 & 0.032 & 0.091 & 0.169 & 0.920 & 0.924 \\
   \bottomrule[1.5pt]
\end{tabular}
\caption{\label{tab:ame_beta_aux} Additional results for from simulation setting 1 involving beta GLLVMs, for response-specific intercepts and overdispersion parameters. Performance was assessed in terms of empirical bias and RMSE, and empirical coverage probability if 95\% Wald intervals. The average bias and RMSE were calculated using a trimming factor of 2 \%. 
}
\end{table*}

\FloatBarrier
\subsection{Setting 2}
\label{appendix:birdsim}

In this second simulationn setting, we compared the performance of EVA, standard VA, and LA for fitting GLLVMs on multivariate data with similar properties to that of Borneo bird data set from \citet{cleary2005associations}. In contrast to the testate amoebae data presented in the main text, the bird data set had a high ratio of the amount of species to the amount of observational sites, with counts of 177 species recorded at 37 sites. From this set, 29 species were considered to be too rare to included in simulation study (observed at less than 3 of the 37 sites). We this omitted these from the simulation study, reducing the total amount of responses (species) to be $m=146$. The bird data set did not contain any environmental covariates. 

The setup of this setting was largely similar to the one used in simulation setting 1 presented in Section $5.1$ of the main text. Specifically, multivariate data were simulated from GLLVMs fitted to the Borneo bird data, where these GLLVMs included only $p = 2$ latent variables. No row effects $\alpha_i$ were included. Using the parameter estimates from these GLLVMs as true parameter values (and hence a true data generation mechanism), we then simulated datasets with differing numbers of responses $m$ while keeping the number of observational units fixed at the same as the original dataset, i.e., $n = 37$. This was accomplished by randomly subsampling the relevant parameter estimates from the fitted GLLVM. We varied the number of species as $m=30, 60, 100, 140$, noting that the full dataset contained 146 species. We simulated 1000 datasets for each value of $m$.

Datasets with four possible response types were generated, following Sections $4.1$ to $4.4$.
\begin{enumerate}
\item Overdispersed counts simulated from a negative binomial GLLVM with log link function fitted to the original Borneo bird count data using the standard VA approach. For each simulated dataset, we then compared negative binomial GLLVMs fitted using EVA to those fitted using standard VA and LA. 

\item Binary responses simulated from a binary GLLVM with either the probit or logit link fitted to a presence-absence version of the original data (formed by setting all positive counts to one, while keeping zero counts at zero). The binary GLLVM was fitted using the standard VA approach (probit link) or the LA approch (logit link). For each simulated dataset, we then compared binary GLLVMs fitted using EVA to those fitted standard VA and/or LA. 

\item Semi-continuous responses simulated from a Tweedie GLLVM with log link fitted to the original count data set using the LA approach. For each simulated dataset, we then compared Tweedie GLLVMs fitted using EVA to those fitted using LA. 

\item Proportions data simulated from a beta GLLVM using the logistic link. As true parameter values for this true model, we used the parmeters of the binary GLLVM with logistic link fitted to the presence-absence version of the original testae amoebae data discussed above. Additionally, the true values of the response-specific dispersion parameters $\phi_j$ were drawn independently from the uniform distribution $\text{Unif}(1,3)$. For each simulated dataset, we then compared beta GLLVMs fitted using EVA to those fitted using LA. \end{enumerate}


\subsubsection{Simulation Results}
 
We discuss results for each of the four response types separately. First, for multivariate overdispersed count data generated from a negative binomial GLLVM, EVA was computationally the most efficient, while LA was easily the slowest method by a clear margin (Table \ref{tab:birds_nb}). We also noted in additional exploration that LA encountered some issues with reaching convergence, although the issue was not as severe as those observed in the analogous situation in simulation setting 1. LA and EVA produced nearly identical results in terms of empirical bias, RMSE and Procrustes errors, while VA was again the most accurate of the three methods. All three methods performed similarly in terms of empirical coverage probability for the response-specific intercept $\beta_{0j}$, but fared poorly in terms of coverage for the dispersion parameters $\phi_j$.

\begin{table*}[tb]
\centering
\begin{tabular}{llrrrrrrrrr}
  \toprule[1.5pt]
 & & & \multicolumn{2}{c}{Bias} & \multicolumn{2}{c}{RMSE} & \multicolumn{2}{c}{Coverage} & \multicolumn{2}{c}{Procrustes} \\
 \cmidrule(lr){4-5} \cmidrule(lr){6-7} \cmidrule(lr){8-9} \cmidrule(lr){10-11}
 $m$&   & Time & $\beta_{0j}$ & $\phi_j$ & $\beta_{0j}$ & $\phi_j$ & $\beta_{0j}$ & $\phi_j$ & LVs & Loadings\\ 
  \midrule
$40$ \quad & EVA & 2.05 & -0.233 & -0.655 & 0.448 & 1.117 & 0.945 & 0.630 & 0.520 & 0.554 \\   
   & VA & 2.58 & 0.069 & 0.158 & 0.311 & 0.917 & 0.865 & 0.822 & 0.486 & 0.553 \\
   & LA & 23.92 & -0.209 & -0.644 & 0.427 & 1.111 & 0.946 & 0.635 & 0.515 & 0.546 \\  \midrule
  $60$ & EVA & 5.95 & -0.246 & -0.531 & 0.424 & 0.975 & 0.962 & 0.685 & 0.235 & 0.459 \\   
   & VA & 7.30 & -0.073 & -0.097 & 0.302 & 0.785 & 0.897 & 0.819 & 0.214 & 0.391 \\
   & LA & 62.93 & -0.242 & -0.527 & 0.419 & 0.973 & 0.962 & 0.687 & 0.232 & 0.449 \\   \midrule
  $100$ & EVA & 12.76 & -0.273 & -0.577 & 0.456 & 1.022 & 0.963 & 0.707 & 0.118 & 0.434 \\  
   & VA & 16.37 & -0.131 & -0.296 & 0.346 & 0.849 & 0.935 & 0.824 & 0.110 & 0.334 \\
   & LA & 139.05 & -0.269 & -0.573 & 0.453 & 1.021 & 0.962 & 0.706 & 0.117 & 0.422 \\  \midrule
  $140$ & EVA & 22.00 & -0.280 & -0.573 & 0.450 & 1.005 & 0.965 & 0.713 & 0.085 & 0.458 \\
   & VA & 25.69 & -0.171 & -0.367 & 0.360 & 0.857 & 0.927 & 0.793 & 0.080 & 0.353 \\ 
   & LA & 210.99 & -0.280 & -0.572 & 0.450 & 1.005 & 0.965 & 0.716 & 0.084 & 0.446 \\ 
   \bottomrule[1.5pt]
\end{tabular}
\caption{\label{tab:birds_nb} Results from simulation setting 2 involving negative binomial GLLVMs. Performance was assessed in terms of computation time, empirical biases and RMSEs of estimated coefficients for the intercept and dispersion parameters, empirical coverage probability if 95\% Wald intervals for the intercept and dispersion parameters, and Procrustes errors of the latent variables and loadings. A trimming factor of $2 \%$ was used when calculating the average biases and RMSEs.
}
\end{table*}

Turning to Bernoulli responses generated from a binary GLLVM with the probit link function, LA was again much slower than both EVA and standard EVA (Table \ref{tab:birds_bin}) in terms of computation. Both EVA and standard VA produced similar results in terms of empirical bias and RMSE, while LA was the most inaccurate. Both EVA and LA had the tendency to produce too narrow Wald intervals, although the undercoverage was only slight. On the other hand, standard VA tended to produce intervals that were a bit too wide leading to overcoverage. In terms of Procrustes errors, VA performed the best overall, while EVA and LA performed poorly in terms of the estimation for the predicted latent variables and loading matrices, respectively.

For the binary GLLVM with logit link case, noting that standard EVA is not available in this scenario, EVA was substantially faster to fit while also producing lower empirical biases and RMSEs (Table \ref{tab:birds_logit}). LA was able to produce better empirical coverage percentages, although the undercoverage but EVA was only slight at larger values of $m$. The performance in terms of the Procrustes errors between the two methods were very similar.

\begin{table*}[htb]
\centering
\begin{tabular}{llcccccc}
  \toprule[1.5pt]
 & & & \multicolumn{1}{c}{Bias} & \multicolumn{1}{c}{RMSE} & \multicolumn{1}{c}{Coverage} & \multicolumn{2}{c}{Procrustes} \\
 \cmidrule(lr){4-4} \cmidrule(lr){5-5} \cmidrule(lr){6-6} \cmidrule(lr){7-8}
 $m$ &   & Time & $\beta_{0j}$ & $\beta_{0j}$ & $\beta_{0j}$ & $u$ & $\lambda$ \\ 
  \midrule
  $40$ & EVA & 0.60 & 0.03 & 0.289 & 0.924 & 0.605 & 0.643 \\
   & VA & 1.11 & 0.04 & 0.227 & 0.950 & 0.494 & 0.540 \\
   & LA & 36.27 & -5.39 & 10.922 & 0.804 & 0.558 & 0.848 \\  \midrule
  $60$ & EVA & 1.35 & 0.00 & 0.265 & 0.946 & 0.405 & 0.541 \\  
   & VA & 2.20 & -0.02 & 0.255 & 0.975 & 0.238 & 0.397 \\ 
   & LA & 63.11 & -2.92 & 9.604 & 0.865 & 0.273 & 0.877 \\  \midrule
  $100$ & EVA & 2.80 & -0.00 & 0.268 & 0.936 & 0.252 & 0.434 \\ 
   & VA & 4.27 & -0.04 & 0.281 & 0.978 & 0.141 & 0.334 \\ 
   & LA & 116.42 & -1.88 & 7.102 & 0.888 & 0.158 & 0.889 \\ \midrule
  $140$ & EVA & 4.72 & -0.02 & 0.251 & 0.939 & 0.153 & 0.369 \\
   & VA & 7.26 & -0.05 & 0.300 & 0.978 & 0.094 & 0.317 \\ 
   & LA & 178.46 & -0.75 & 4.928 & 0.913 & 0.105 & 0.892 \\
   \bottomrule[1.5pt]
\end{tabular}
\caption{\label{tab:birds_bin} Results from simulation setting 2 involving binary GLLVMs with the probit link function. Performance was assessed in terms of computation time, empirical biases and RMSEs of estimated coefficients for the intercept, empirical coverage probability if 95\% Wald intervals for the intercept, and Procrustes errors of the latent variables and loadings. A trimming factor of $2 \%$ was used when calculating the average biases and RMSEs.
}
\end{table*}

\begin{table*}[htb]
\centering
\begin{tabular}{llcccccc}
  \toprule[1.5pt]
 & &             & \multicolumn{1}{c}{Bias} & \multicolumn{1}{c}{RMSE} & \multicolumn{1}{c}{Coverage} & \multicolumn{2}{c}{Procrustes} \\
 \cmidrule(lr){4-4} \cmidrule(lr){5-5} \cmidrule(lr){6-6} \cmidrule(lr){7-8}
 $m$ &   & Time & $\beta_{0j}$ & $\beta_{0j}$ & $\beta_{0j}$ & $u$ & $\lambda$ \\ 
  \midrule
$40$ & EVA & 0.34 & 2.32 & 3.850 & 0.897 & 0.563 & 0.902 \\ 
   & LA & 22.29 & -6.92 & 25.914 & 0.780 & 0.553 & 0.866 \\   \midrule
  $60$ & EVA & 0.81 & -0.25 & 3.592 & 0.921 & 0.275 & 0.840 \\   
   & LA & 25.18 & -4.12 & 14.902 & 0.928 & 0.275 & 0.833 \\  \midrule
  $100$ & EVA & 1.78 & -1.43 & 3.125 & 0.926 & 0.153 & 0.869 \\ 
   & LA & 42.99 & -3.27 & 11.352 & 0.943 & 0.160 & 0.857 \\ \midrule
  $140$ & EVA & 3.08 & -1.83 & 3.409 & 0.922 & 0.091 & 0.687 \\ 
   & LA & 69.71 & -2.29 & 8.198 & 0.946 & 0.099 & 0.684 \\
   \bottomrule[1.5pt]
\end{tabular}
\caption{\label{tab:birds_logit} Results from simulation setting 2 involving binary GLLVMs with the logit link function. Performance was assessed in terms of computation time, empirical biases and RMSEs of estimated coefficients for the intercept, empirical coverage probability if 95\% Wald intervals for the intercept, and Procrustes errors of the latent variables and loadings. A trimming factor of $2 \%$ was used when calculating the average biases and RMSEs.
}
\end{table*}

Next, for semi-continuous data generated from a Tweedie GLLVM, noting that standard EVA is not available in this setting, the performance between EVA and LA was very similarly for all performance metrics except computation speed, where EVA was considerably faster to fit (Table \ref{tab:birds_tweedie}). Both EVA and LA produced too narrow confidence intervals regarding the dispersion parameters $\phi_j$.

\begin{table*}[htb]
\centering
\begin{tabular}{llrcccccccc}
  \toprule[1.5pt]
 & & & \multicolumn{2}{c}{Bias} & \multicolumn{2}{c}{RMSE} & \multicolumn{2}{c}{Coverage} & \multicolumn{2}{c}{Procrustes} \\
 \cmidrule(lr){4-5} \cmidrule(lr){6-7} \cmidrule(lr){8-9} \cmidrule(lr){10-11}
 $m$&   & Time & $\beta_{0j}$ & $\phi_j$ & $\beta_{0j}$ & $\phi_j$ & $\beta_{0j}$ & $\phi_j$ & LVs & Loadings\\ 
  \midrule
$40$ & EVA & 4.54  & -0.153 & -0.050 & 0.418 & 0.259 & 0.961 & 0.905 & 0.420 & 0.518 \\ 
   & LA & 22.54  & -0.125 & -0.049 & 0.397 & 0.258 & 0.960 & 0.905 & 0.417 & 0.508 \\   \midrule
  $60$ & EVA & 11.72  & -0.178 & -0.039 & 0.397 & 0.252 & 0.970 & 0.906 & 0.186 & 0.399 \\ 
   & LA & 72.24  & -0.167 & -0.038 & 0.388 & 0.252 & 0.970 & 0.906 & 0.185 & 0.386 \\   \midrule
  $100$ & EVA & 30.28  & -0.198 & -0.039 & 0.460 & 0.261 & 0.969 & 0.904 & 0.102 & 0.424 \\
   & LA & 201.03  & -0.190 & -0.039 & 0.454 & 0.261 & 0.968 & 0.904 & 0.102 & 0.416 \\  \midrule
  $140$ & EVA & 48.70  & -0.212 & -0.039 & 0.437 & 0.256 & 0.971 & 0.908 & 0.070 & 0.430 \\ 
   & LA & 337.38  & -0.207 & -0.039 & 0.434 & 0.256 & 0.971 & 0.908 & 0.069 & 0.421 \\
   \bottomrule[1.5pt]
\end{tabular}
\caption{\label{tab:birds_tweedie} Results from simulation setting 2 involving Tweedie GLLVMs. Performance was assessed in terms of computation time, empirical biases and RMSEs of estimated coefficients for the intercept and dispersion parameters, empirical coverage probability if 95\% Wald intervals for the intercept and dispersion parameters, and Procrustes errors of the latent variables and loadings. A trimming factor of $2 \%$ was used when calculating the average biases and RMSEs.
}
\end{table*}

Finally, for proportions data generated from a beta GLLVM assuming the logit link function, and noting that standard EVA is not available in this setting, LA took more than ten times longer to fit the GLLVM compared to EVA for the two large values of $m$ considered (Table \ref{tab:birds_beta}). On the other hand, LA did produce more accurate estimates of the response-specific intercepts overall,, while empirical coverage probabilities for both fairly poor. Both methods produced similar results in terms of the Procrustes errors for the predicted latent variables and loading matrices.

\begin{table*}[tb]
\centering
\begin{tabular}{llrcccccccc}
  \toprule[1.5pt]
 & & & \multicolumn{2}{c}{Bias} & \multicolumn{2}{c}{RMSE} & \multicolumn{2}{c}{Coverage} & \multicolumn{2}{c}{Procrustes} \\
 \cmidrule(lr){4-5} \cmidrule(lr){6-7} \cmidrule(lr){8-9} \cmidrule(lr){10-11}
 $m$&   & Time & $\beta_{0j}$ & $\phi_j$ & $\beta_{0j}$ & $\phi_j$ & $\beta_{0j}$ & $\phi_j$ & LVs & Loadings\\ 
  \midrule
$40$ \quad & EVA & 8.57 & 2.063 & 0.403 & 2.300 & 0.758 & 0.920 & 0.923 & 0.273 & 0.939 \\ 
   & LA & 42.87 & 2.028 & 0.283 & 2.192 & 0.579 & 0.926 & 0.927 & 0.210 & 0.946 \\  \midrule
  $60$ & EVA & 22.22 & -0.006 & 0.251 & 2.205 & 0.514 & 0.927 & 0.929 & 0.097 & 0.979 \\ 
   & LA & 140.19 & -0.006 & 0.254 & 2.192 & 0.506 & 0.928 & 0.930 & 0.092 & 0.979 \\    \midrule
  $100$ & EVA & 39.81 & 0.356 & 0.362 & 0.848 & 0.641 & 0.671 & 0.896 & 0.088 & 0.986 \\
   & LA & 464.10 & 0.174 & 0.393 & 0.611 & 0.647 & 0.825 & 0.897 & 0.066 & 0.985 \\  \midrule
  $140$ & EVA & 73.47 & 0.309 & 0.313 & 0.955 & 0.632 & 0.653 & 0.878 & 0.103 & 0.984 \\
   & LA & 816.78 & 0.191 & 0.304 & 0.845 & 0.606 & 0.755 & 0.869 & 0.085 & 0.983 \\ 
   \bottomrule[1.5pt]
\end{tabular}
\caption{\label{tab:birds_beta} Results from simulation setting 2 involving beta GLLVMs. Performance was assessed in terms of computation time, empirical biases and RMSEs of estimated coefficients for the intercept and dispersion parameters, empirical coverage probability if 95\% Wald intervals for the intercept and dispersion parameters, and Procrustes errors of the latent variables and loadings. A trimming factor of $2 \%$ was used when calculating the average biases and RMSEs.
}
\end{table*}

\FloatBarrier
\section{Additional results for application to testate amoebae data}
\label{appendix:example}

\begin{figure}
    \centering
    \includegraphics[width=\textwidth]{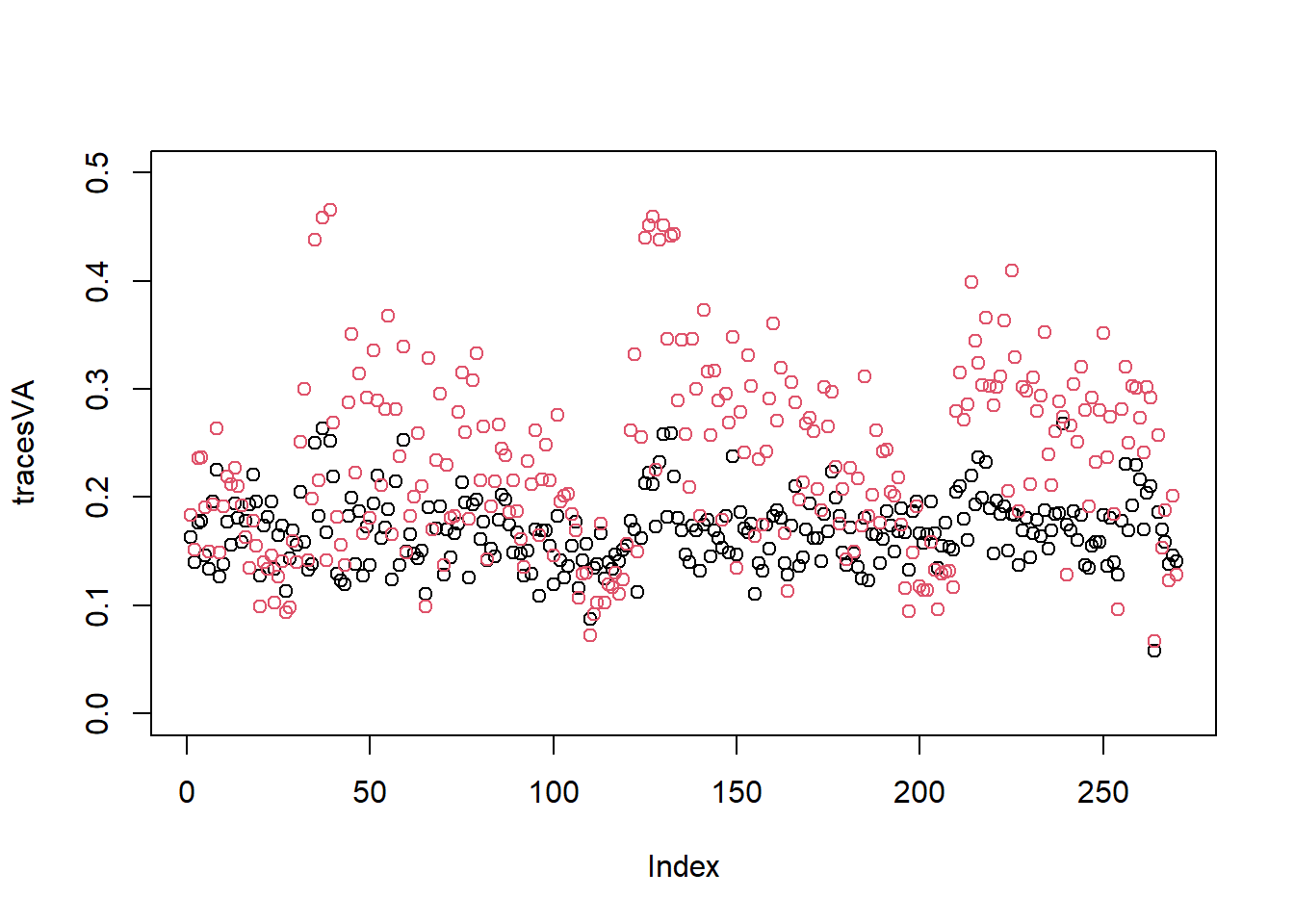}
    \caption{Traces of the estimated variational covariance matrices $\bm{\hat A_i}$ from the model containing water pH, temperature, and land use. The black dots correspond to VA, while the red dots correspond to EVA.}
    \label{fig:varicov_traces}
\end{figure}


\begin{figure}[ht]
    \centering
    \includegraphics[height=0.4\textheight]{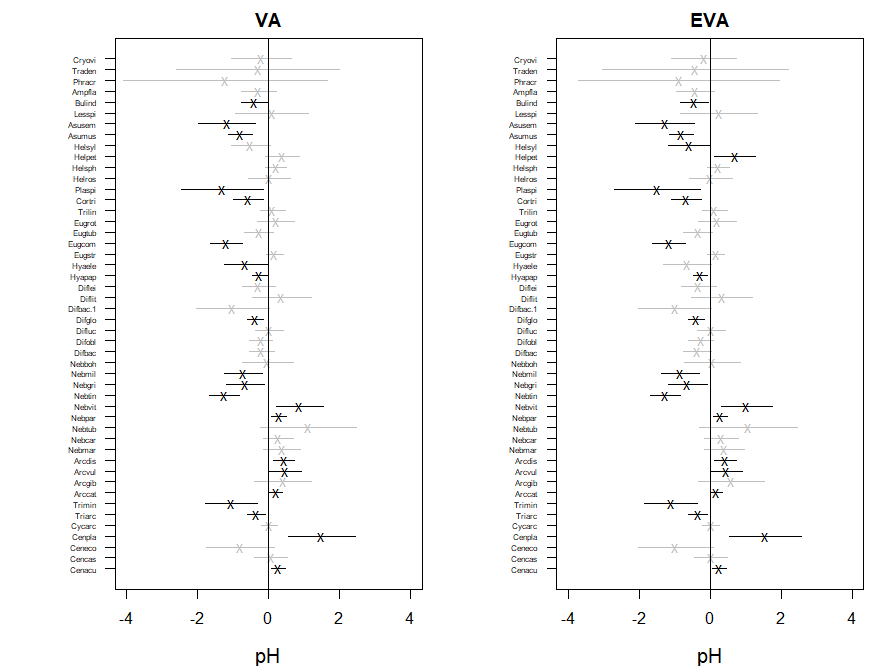}
    \includegraphics[height=0.4\textheight]{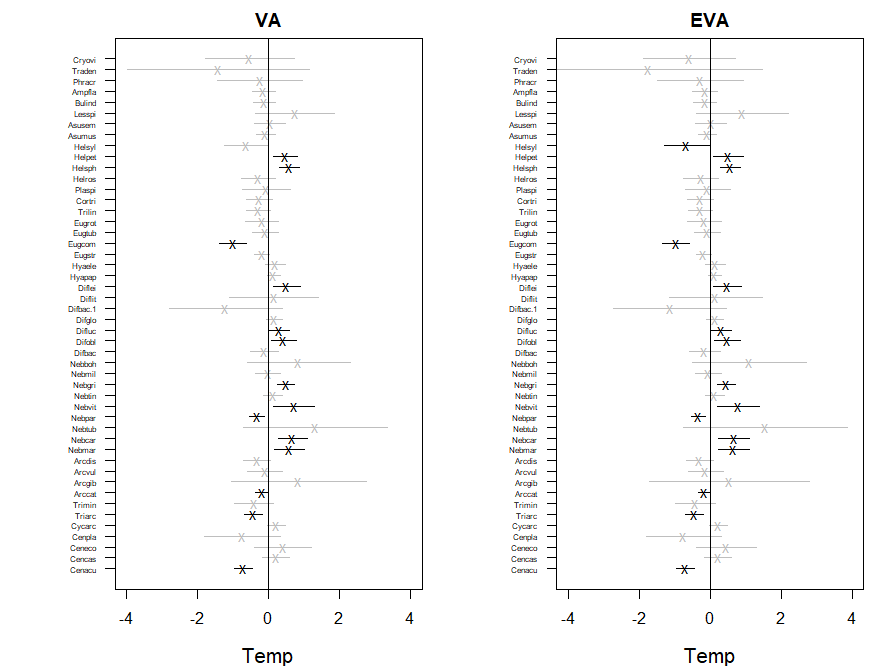}
    \caption{Coefficient plots containing the point estimates and $95 \%$ Wald intervals for the effect of water pH (top row) and temperature (bottom row) for the $m=48$ species in the testae amoebae data. The plot in the left column corresponds to negative binomial GLLVMs fitted using standard VA, while the plot in the right column corresponds to those fitted EVA.}
    \label{fig:coefplot_pH}
\end{figure}

\begin{figure}[ht]
    \centering
    \includegraphics[height=0.4\textheight]{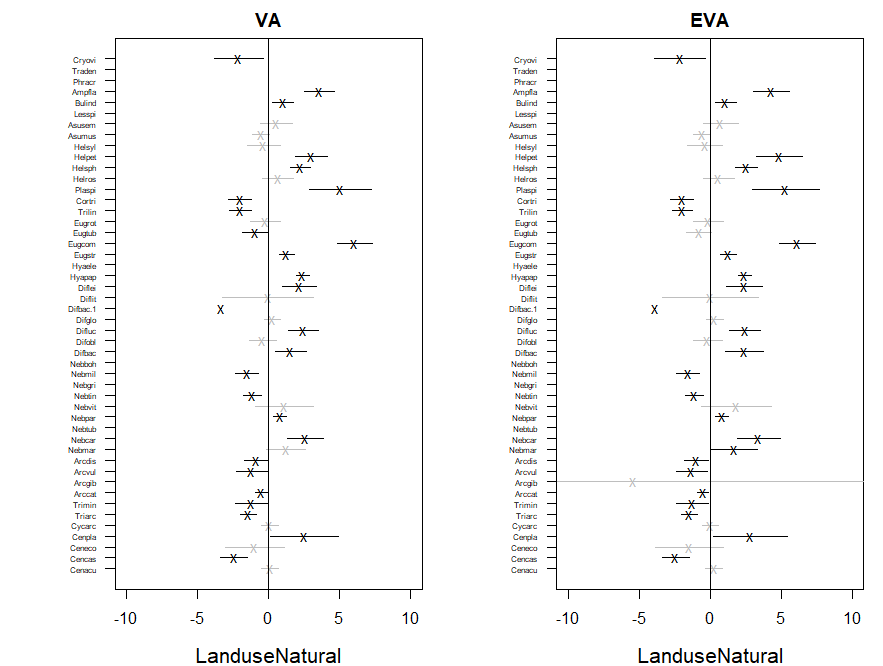}
    \includegraphics[height=0.4\textheight]{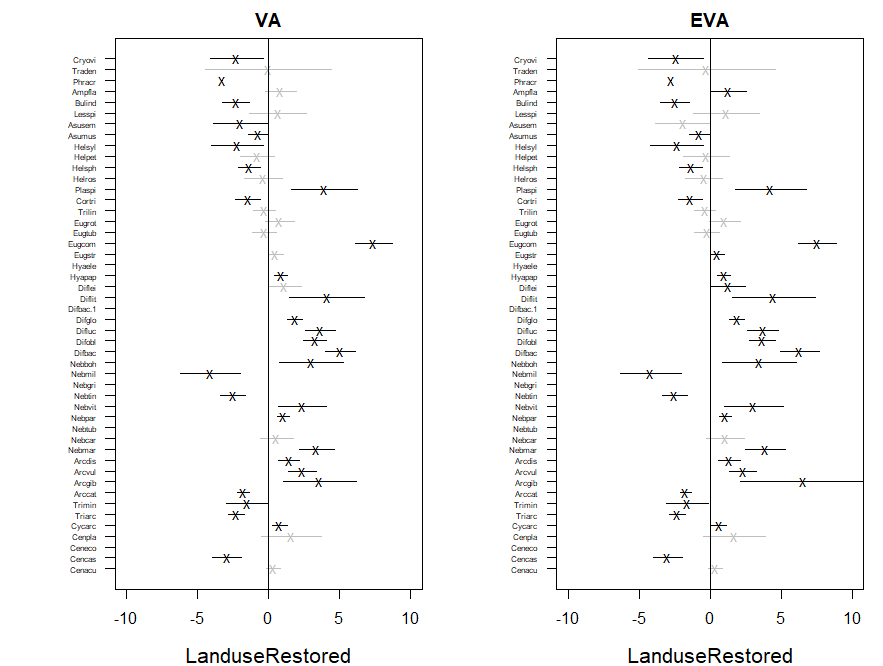}
    \caption{Coefficient plots containing the point estimates and $95 \%$ Wald intervals for the effects of natural (top row) and restored (bottom row) land use types for the $m=48$ species in the testae amoebae data. The plot in the left column corresponds to negative binomial GLLVMs fitted using standard VA, while the plot in the right column corresponds to those fitted EVA.}
    \label{fig:coefplot_landuse}
\end{figure}






\end{document}